\documentclass[aps,nofootinbib,preprint,tightenlines]{revtex4}

\usepackage{graphicx}
\usepackage{amsmath}
\usepackage{bm} 

\newcommand{\beq}{\begin{eqnarray}}
\newcommand{\eeq}{\end{eqnarray}}
\newcommand{\bqa}{\begin{eqnarray}}
\newcommand{\eqa}{\end{eqnarray}}

\begin{document}


\preprint{INT-PUB-11-007}
%
\title{%
Electric properties of the Beryllium-11 system in Halo EFT
}
\author{H.-W. Hammer}\email{hammer@hiskp.uni-bonn.de}
\affiliation{Helmholtz-Institut f\"ur Strahlen- und Kernphysik (Theorie)
and Bethe Center for Theoretical Physics,
 Universit\"at Bonn, 53115 Bonn, Germany}
\author{D.~R.~Phillips}\email{phillips@phy.ohiou.edu}
\affiliation{Department of Physics and Astronomy, Ohio University, Athens, Ohio 45701, USA}
\affiliation{Helmholtz-Institut f\"ur Strahlen- und Kernphysik (Theorie)
and Bethe Center for Theoretical Physics,
 Universit\"at Bonn, 53115 Bonn, Germany}
\begin{abstract}
We compute E1 transitions and electric radii in the Beryllium-11 
nucleus using an effective field theory that exploits the separation of 
scales in this halo system. We fix the leading-order parameters of the EFT 
from measured data on the 1/2$^+$ and 1/2$^-$ levels in ${}^{11}$Be and the 
B(E1) strength for the transition between them. We then obtain predictions 
for the B(E1) strength for Coulomb dissociation of the ${}^{11}$Be nucleus 
to the continuum. We also compute the charge radii of the 1/2$^+$ and 1/2$^-$ 
states. Agreement with experiment within the expected accuracy of a 
leading-order computation in this EFT is obtained. We also discuss how 
next-to-leading-order (NLO) corrections involving both s-wave and p-wave 
${}^{10}$Be-neutron interactions affect our results, and display the NLO 
predictions for quantities which are free of additional short-distance 
operators at this order. Information on neutron-${}^{10}$Be scattering in 
the relevant channels is inferred. 
\end{abstract} 
\maketitle
%
%
%
\section{Introduction}
\label{intro}


The first excitation of the Beryllium-10 nucleus is 3.4 MeV above the ground 
state, and that ground state has 
spin and parity quantum numbers $J^P=0^+$. Meanwhile, 
the Beryllium-11 nucleus has a $1/2^+$ state whose neutron separation energy 
is 500 keV, and a $1/2^-$ state whose neutron separation energy is 180 keV~\cite{AjK90}. 
The shallowness of these two states of ${}^{11}$Be compared to the bound 
states of ${}^{10}$Be suggests that they have significant components in 
which a loosely-bound neutron orbits a ${}^{10}$Be core. In this ``one-neutron
halo" picture the $1/2^+$ is predominantly an s-wave bound state, while the 
$1/2^-$ is predominantly a relative p-wave between the neutron and the core. 
In this paper, we discuss efforts to use effective field theory (EFT) to 
systematically implement such a halo picture of the ${}^{11}$Be nucleus. 
 
This halo viewpoint is reinforced by the fact that the scattering volume 
of n-${}^{10}$Be scattering in the $l=1$, $J=1/2$ channel has been determined 
to be~\cite{TB04}
\begin{equation}
a_1=(457 \pm 67)~{\rm fm}^3.
\label{eq:pwavescattlength}
\end{equation}
The corresponding length scale of order 8 fm is large compared to the 
natural length-scale of core-neutron interactions, which is $\approx 
2$--$3$ fm. 

The datum (\ref{eq:pwavescattlength}), together with the information on 
the bound-state energies in the ${}^{10}$Be and ${}^{11}$Be systems, 
helps us to estimate the expansion parameter in our Halo EFT. This is 
the binding energy of the halo nucleus, as compared to the energy required 
to excite the core, i.e. $B_{\rm lo}/B_{\rm hi} \approx 1/6$. Converting 
this to an estimate of the different distance scales involved, we infer 
that a majority of the probability density of ${}^{11}$Be occupies a region 
outside the ${}^{10}$Be core: $R_{\rm core}/R_{\rm halo}$ $\approx 0.4$, 
which is consistent with the ratio implied by the numbers in the previous 
paragraph. This ratio of distance scales is the formal expansion parameter 
for the EFT, and since it is not particularly small, leading-order 
calculations are only a first step. We therefore present calculations up to 
next-to-leading order for several quantities, in order to confirm that 
the series is converging as expected. 

In particular we apply this EFT to electromagnetic reactions in the ${}^{11}$Be 
system. The B(E1)($1/2^+ \rightarrow 1/2^-$) transition has recently been 
measured to be 
\begin{equation}
{\rm B(E1)}(1/2^+ \rightarrow 1/2^-)=0.105(12)~e^2~{\rm fm}^2
\label{eq:CEXBE1}
\end{equation}
using intermediate-energy Coulomb excitation~\cite{Su07}. 
This is consistent with the older number 
\begin{equation}
{\rm B(E1)}(1/2^+ \rightarrow 1/2^-)=0.116(12)~e^2~{\rm fm}^2
\label{eq:lifeBE1}
\end{equation}
from lifetime measurements~\cite{Mi83}. There are also two recent data 
sets on the Coulomb-induced breakup of the ${}^{11}$Be 
nucleus~\cite{Pa03,Fu04} (see also Ref.~\cite{An94}). Both experiments 
extracted the excitation function $d$B(E1)$/dE$ as a function of the 
energy of the outgoing neutron $E$. For low neutron energies this 
excitation function is affected by the final-state interaction in the 
p-waves, and can be predicted in the halo picture~\cite{TB04}. Ref.~\cite{Pa03} also
extracted a neutron radius for the ground state of ${}^{11}$Be from their data:
\begin{equation}
\langle r^2 \rangle^{1/2}=5.7(4)~{\rm fm}.
\end{equation}
This is consistent with the recent atomic-physics measurement of 
the ${}^{11}$Be charge radius~\cite{No09}:
\begin{equation}
\langle r_E^2 \rangle_{{}^{11}{\rm Be}}^{1/2}=2.463(16)~{\rm fm}.
\label{eq:atphysnumber}
\end{equation}

All of these measurements can be addressed within the Halo EFT we will 
use here. In this theory  the s- and p-wave states of the Beryllium-11 
nucleus are generated by core-neutron contact interactions. The theory 
does not get the interior part of the nuclear wave function correct, but, 
by construction, it reproduces the correct asymptotics of the wave functions 
of these states:
\begin{eqnarray}
u_0(r)&=&A_0 \exp(-\gamma_0 r)\,,\nonumber\\ 
u_1(r)&=&A_1 \exp(-\gamma_1 r) \left(1 + \frac{1}{\gamma_1 r} \right)\,,
\label{eq:LOwfs}
\end{eqnarray}
for the $500$ keV and 180 keV states, respectively. 
The quantities $\gamma_0$ and $\gamma_1$ are determined by the neutron 
separation energies of the states in question. At leading order (LO) in 
the expansion the Asymptotic Normalization Coefficients (ANCs) 
$A_0$ and $A_1$ are fixed. (In the case of the p-wave 
this is related to the theorem discussed in Ref.~\cite{Lee}.) 
However, at next-to-leading order
$A_0$ and $A_1$ become, in essence, free parameters of the theory, and
must themselves be extracted from data.

Halo EFT is well-suited for this task. It is not intended to
compete with {\it ab initio} calculations of this 
halo nucleus (see, e.g.~\cite{Fo05,Fo09}) or of ${}^{10}$Be-n 
scattering~\cite{QN08}, or with microscopic descriptions of the ${}^{11}$Be
E1 strength (see, e.g.~\cite{Mi83}). Instead, Halo EFT is complementary 
to such approaches, since it takes $A_0$, $A_1$, $\gamma_0$, 
and $\gamma_1$ as input, rather than seeking to predict them via a detailed 
description of the $A=11$ system. 
The EFT's goal is to ensure that the 
long-distance properties of the halo are correctly taken care of, and then 
to elucidate the relationships between different observables in  the 
${}^{10}$Be-n
system that result. Such correlations flow directly from the existence of 
the shallow $1/2^+$ and $1/2^-$ bound
states in this system. 
In particular, below we will show how $A_0$ and $A_1$
are correlated with neutron-$^{10}$Be scattering observables, as well as with 
${\rm B(E1)}(1/2^+ \rightarrow 1/2^-)$ and
the Coulomb dissociation data.
This, in turn, demonstrates how---and with what accuracy---the ANCs can be 
inferred from these experimental quantities.

A preliminary version of these findings appeared in Ref.~\cite{HP10}. 
The presentation here corrects and expands upon that earlier work. 
A related study of radiative neutron capture on Lithium-7 was recently
carried out in~\cite{Rupak:2011nk}. The mechanism for the cancellation 
of divergences in the s-to-p transition in this reaction and in
the Beryllium-11 system is the same. 

\section{Halo EFT for Beryllium-11}

We use the ``Halo EFT" developed in Refs.~\cite{Be02,Bd03} to calculate 
the properties of the Beryllium-11 nucleus. 
The degrees of freedom in our Halo EFT treatment are the ${}^{10}$Be core 
and the neutron. The EFT expansion in this case is an expansion in powers 
of $\omega/B_{\rm high}$. Here $B_{\rm high}$ is, e.g. the excitation 
energy of states in ${}^{10}$Be, and so is of order a few MeV, and $\omega$ 
is the energy of the photon exciting the electromagnetic transition of 
interest. 

\subsection{Lagrangian: strong sector}

In our calculation, the $^{10}$Be core and the neutron are represented by 
the bosonic field $c$ and a spinor field $n$, respectively.
We include the strong s-wave and p-wave interactions that lead to the 
shallow bound states in the ${}^{11}$Be system through the incorporation 
of additional fields, which encode physics of these states. Therefore the 
$1/2^+$ state is constructed as a spinor field, $\sigma_s$. In contrast, 
the field representing the $1/2^-$ state is a pseudo-spinor, i.e. it is 
parity odd. We denote it as $\pi_s$. Its behavior under parity restricts 
the types of couplings which can appear in the Lagrangian. In particular, 
only combinations of nucleon and core fields with an odd number of 
derivatives can couple to $\pi$. Moreover, when constructing these 
operators we must decompose them into their irreducible representations 
under the rotation group. The portion that couples to $\pi_s$ is then 
just the $J=1/2$ part. For example, if we construct $n_\beta (i 
\stackrel{\leftrightarrow}{\nabla}_j) c$ where 
$\stackrel{\leftrightarrow}{\nabla}_j=(\stackrel{\leftarrow}{\nabla}
-\stackrel{\rightarrow}{\nabla})_j$,
this combination has both $J=1/2$ 
and $J=3/2$ parts. We project out the $J=1/2$ part by defining:
\begin{equation}
[n (i \stackrel{\leftrightarrow}{\nabla}) c]_{{1 \over 2}, s}
=\sum_{\beta j}\left({1 \over 2} \beta 1 j\right|\left.
\left({1 \over 2} 1\right){1 \over 2} s \right) n_\beta \, (i  
\stackrel{\leftrightarrow}{\nabla}_j) \, c\,,
\end{equation}
where $(j_1 m_1 j_2 m_2 | (j_1 j_2) J M)$ is the Clebsch-Gordan
coefficient to couple $j_1$ and $j_2$ to $J$.

With the masses of the neutron  and the core are denoted by $m$ and $M$, 
respectively, and $M_{nc}=(M+m)$ is the total mass of the n-$^{10}$Be system, 
we then have: 
\begin{eqnarray}
{\cal L}&=&c^\dagger \left(i \partial_t + \frac{\nabla^2}{2 M}\right)c 
+ n^\dagger \left(i \partial_t + \frac{\nabla^2}{2 m}\right)n \nonumber\\
&&+ \sigma_s^\dagger \left[\eta_0 \left(i \partial_t + 
\frac{\nabla^2}{2M_{nc}}\right) + \Delta_0\right]\sigma_s \nonumber 
+ \pi_s^\dagger \left[\eta_1 \left(i \partial_t + 
\frac{\nabla^2}{2M_{nc}}\right) + \Delta_1\right] \pi_s\nonumber \\
&& - g_0  \left[c^\dagger n_s^\dagger \sigma_s + \sigma_s^\dagger 
n_s c\right]- \frac{g_1}{2} \left( \pi_s^\dagger \left[n 
(i\stackrel{\leftrightarrow}{\nabla}) c\right]_{\frac{1}{2},s}
+ \left[c^\dagger (i\stackrel{\leftrightarrow}{\nabla}) n^\dagger
\right]_{\frac{1}{2},s} \pi_s\right) \nonumber \\
&&- \frac{g_1}{2} \frac{M-m}{M_{nc}}\left(\pi_s^\dagger 
\left[i\stackrel{\rightarrow}{\nabla} (nc)\right]_{\frac{1}{2},s}
- \left[i\stackrel{\rightarrow}{\nabla} (n^\dagger c^\dagger)
\right]_{\frac{1}{2},s} \pi_s\right) + \ldots\,,
\label{eq:HEFT}
\end{eqnarray}
where we adopt the convention that repeated spin indices are summed. 
Note that the last line of Eq.~(\ref{eq:HEFT}) represents an additional 
p-wave interaction necessary to 
maintain Galilean invariance. It is required because the $n$ and $c$ 
fields have different masses.
The dots represent higher-order interactions not considered here. 
One such interaction involves the
$J=3/2$ part of the $[n (i \stackrel{\leftrightarrow}{\nabla}) c]$ operator:
\begin{eqnarray}
{\cal L}_{3/2}&=&-\frac{C^{(3/2)}}{4}[n (i \stackrel{\leftrightarrow}{\nabla})
c]^\dagger_{\frac{3}{2},\beta}[n (i \stackrel{\leftrightarrow}{\nabla}) 
c]_{\frac{3}{2},\beta}\nonumber\\
&=& -\frac{C^{(3/2)}}{4} \sum_{\alpha s_1 j s_2 k} \left(\frac{1}{2} 
s_1 1 j\right|\left. \left(\frac{1}{2} 1\right) \frac{3}{2} \alpha\right) 
\left(\frac{1}{2} s_2 1 k\right|\left. \left(\frac{1}{2} 1\right) 
\frac{3}{2} \alpha\right)  (c^\dagger (i 
\stackrel{\leftrightarrow}{\nabla}_j) n_{s_1}^\dagger)(n_{s_2}  (i 
\stackrel{\leftrightarrow}{\nabla}_k) c)\,.
\nonumber\\
\end{eqnarray}

As we shall discuss in the next section, this interaction is assumed 
to be natural in our power counting, in contrast to the interactions 
mediated by $\pi$ and $\sigma$ fields, which are enhanced. 

\subsection{s-wave ${}^{10}$Be-neutron interactions}

\label{sec-swaves}

In order to treat the shallow s-wave state in the ${}^{10}$Be-neutron 
system we adopt the counting that has been successfully developed to treat 
shallow s-wave states in the nucleon-nucleon 
system~\cite{vK99,Ka98A,Ka98B,Ge98,Bi99}. 
In leading order, the $\sigma$ field is static and its bare propagator
is simply $1/\Delta_0$. Dividing out the leading term, the one-loop 
correction to the bare propagator
is $g_0^2/\Delta_0$ times an $n c$ bubble which has a  typical 
size of order $1/R_{\rm halo}$~\footnote{In a suitable regularization 
scheme, e.g. power-law divergence subtraction~\cite{Ka98A,Ka98B}, this 
is true for both the real and imaginary parts of the loops.}.
Since  $g_0^2/\Delta_0\sim R_{\rm halo}$, this correction
is of order one. Consequently, the $nc$ loops must be
resummed when computing the  full $\sigma$ propagator.

\begin{figure}[!t]
\centerline{\includegraphics[width=0.75\columnwidth]{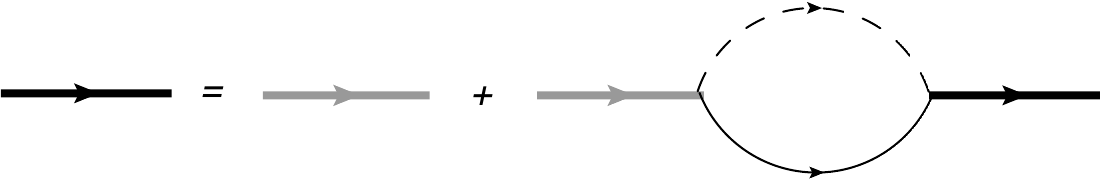}}
\caption{Diagrammatic representation of the Dyson equation for the 
dressed $\sigma$ propagator
representing the s-wave ${}^{10}$Be-n bound state in the theory. 
Here and below the dashed line indicates 
the field for the ${}^{10}$Be core, and the thin solid line is the 
neutron. The thick grey line is the 
bare $\sigma$ propagator, and the thick black line is the dressed 
$\sigma$ propagator.}
\label{fig:sigmadressing}       
\end{figure}

This can be achieved through the Dyson equation shown in 
Fig.~\ref{fig:sigmadressing}, which leads to:
\begin{equation}
D_\sigma(p)=
\frac{1}{\Delta_0 + \eta_0[p_0 - {\bf p}^2/(2 M_{nc})+i\epsilon] 
- \Sigma_\sigma(p)}\,,
\label{eq:dressprop}
\end{equation}
with $\Sigma_\sigma(p)$ the one-loop self-energy for the $\sigma$ field. 

This one-loop self-energy is calculated as:
\begin{equation}
\Sigma_\sigma(p)=-\frac{g_0^2 m_R}{2 \pi} \left[i\sqrt{2 m_R 
\left(p_0 - \frac{{\bf p}^2}{2 M_{nc}}\right)}+\mu\right]\,,
\label{eq:Sigsig}
\end{equation}
when computed in power-law divergence subtraction (PDS) with a 
scale $\mu$~\cite{Ka98A,Ka98B}. Here we have introduced the 
reduced mass of the neutron-core system:
\begin{equation}
m_R=\frac{m M}{m + M}\,,
\end{equation}
and the limit $\epsilon\to 0^+$ in the end is understood.

\begin{figure}[!t]
\centerline{\includegraphics[width=0.5\columnwidth]{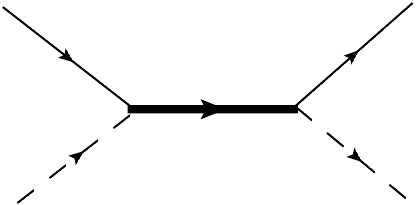}}
\caption{The Feynman diagram relating the dressed $\sigma$ propagator
to the s-wave neutron-core scattering amplitude.}
\label{fig:swavetmatrix}       
\end{figure}

Substituting Eq.~(\ref{eq:Sigsig}) into Eq.~(\ref{eq:dressprop}), we can 
set the parameters $g_0$ and $\Delta_0$ by computing the s-wave neutron-core 
scattering amplitude in the theory defined by Eq.~(\ref{eq:HEFT}) (see 
Fig.~\ref{fig:swavetmatrix}):
\begin{equation}
t_0(E)=g_0^2 D_\sigma(E,{\bf 0})\,,
\label{eq:t0sigma}
\end{equation}
in the two-body center-of-mass frame with $E=k^2/(2 m_R)$. This is then
matched to the effective-range expansion in this channel:
\begin{equation}
t_0(E)=\frac{2 \pi}{m_R}\frac{1}{1/a_0 - \frac{1}{2} r_0 k^2 + i k}\,,
\label{eq:t0ERE}
\end{equation}
producing
\begin{equation}
D_\sigma(p)=\frac{2 \pi \gamma_0}{m_R^2 g_0^2} \frac{1}{1 - r_0 \gamma_0} 
\frac{1}{p_0 - \frac{{\bf p}^2}{2 M_{nc}} + B_0} + R_\sigma(p)\,,
\label{eq:Dsigmadressed}
\end{equation}
where $R_\sigma(p)$ is regular at the pole $p_0 - \frac{{\bf p}^2}{2 M_{nc}}
=-B_0$. 
In Eq.~(\ref{eq:Dsigmadressed}), the position of the pole is determined 
by the binding energy $B_0=\gamma_0^2/(2 m_R)$, and $\gamma_0$ is the 
positive root of the equation:
\begin{equation}
\frac{1}{a_0} + \frac{1}{2} r_0 \gamma_0^2 - \gamma_0=0\,.
\label{eq:gamma0}
\end{equation}
The wave-function renormalization for the dressed $\sigma$ propagator
can be read off as the residue of the pole in Eq.~(\ref{eq:Dsigmadressed}):
\begin{equation}
Z_\sigma=\frac{2\pi\gamma_0}{m_R^2 g_0^2} \left(1-\gamma_0 r_0\right)^{-1}\,.
\end{equation}
This result is valid to NLO in the expansion in $R_{\rm core}/R_{\rm halo} 
\sim r_0 \gamma_0$. It yields a wave function (\ref{eq:LOwfs}) with 
\begin{equation}
A_0=\sqrt{\frac{2 \gamma_0}{1-\gamma_0 r_0}}\,.
\label{eq:A0NLO}
\end{equation}

\subsection{p-wave ${}^{10}$Be-neutron interactions}
\label{sec:p-wave-int}

\begin{figure}[!t]
\centerline{\includegraphics[width=0.75\columnwidth]{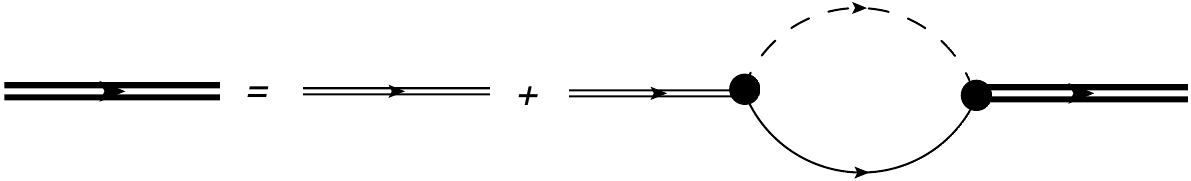}}
\caption{
Diagrammatic representation of the Dyson equation for the dressed $\pi$ 
propagator
representing the p-wave ${}^{10}$Be-n bound state in the theory.
Once again, the dashed line indicates the field for the ${}^{10}$Be core, 
and the thin solid line is the neutron. The thin double line is the bare 
$\pi$ propagator, and the thick double line is the dressed $\pi$ propagator.}
\label{fig:pidressing}       
\end{figure}

We proceed similarly for the p-wave state. The propagator, 
$D_\pi(p)_{ss'}=D_\pi(p) \delta_{ss'}$,
for this state obeys the Dyson equation depicted in Fig.~\ref{fig:pidressing}.
Rather than computing the self-energy of the $\pi$ field directly it is easier
to compute the self-energy for a field $n_\beta (i 
\stackrel{\leftrightarrow}{\nabla}_j) c$ and then couple the neutron spin and 
the relative momentum in the appropriate way to project out the $J=1/2$ piece 
of the result. Hence we now consider the one-loop self-energy, 
$\Sigma_\pi(p)_{ij,\beta \alpha}$ for such a p-wave field. We first observe:
\begin{equation}
\Sigma_\pi(p)_{ij,\alpha \beta}=\delta_{ij} \delta_{\alpha \beta} \Sigma(p)\,.
\end{equation}
The scalar function:
\begin{equation}
\Sigma(p)=
-\frac{m_R g_1^2}{6 \pi} 2m_R \left(p_0 - \frac{{\bf p}^2}{2 M_{nc}}\right) 
\left[i \sqrt{2m_R \left(p_0 - \frac{{\bf p}^2}{2 M_{nc}}+i\epsilon\right)}
+\mu \right],
\end{equation}
where the PDS scheme has been employed and momentum traces have 
been performed 
in three dimensions. 
From this we can construct a self-energy for transitions from the $\pi$-field 
state $s$ to the $\pi$-field state $s'$:
\begin{equation}
{\Sigma_{\pi}}_{s's}(p)=\sum_{\beta j}  \left(\frac{1}{2} \beta 1 j\right|
\left.\left(\frac{1}{2} 1\right) \frac{1}{2} s\right) \left(\frac{1}{2} 
\beta 1 j\right|\left.\left(\frac{1}{2} 1\right) \frac{1}{2} s'\right) 
\Sigma(p)
\end{equation}
since $\Sigma$ is independent of $\beta$ and $j$ 
we can use completeness of the 
Clebsch-Gordon coefficients to show that $\Sigma$ is diagonal in $s'$ and 
$s$, i.e. ${\Sigma_{\pi}}_{s's}(p)=\delta_{s's} \Sigma(p)$. It follows 
that 
$D_\pi(p)$ takes the form:
\begin{equation}
D_\pi(p)=
\frac {1}{\Delta_1 + \eta_1[p_0 - {\bf p}^2/(2 M_{nc})] - \Sigma(p)}\,.
\end{equation}

We note that since the self-energy loop is cubically divergent both 
parameters, $\Delta_1$ and $g_1$ are mandatory for renormalization at 
LO \cite{Be02}. 
This time we are interested in the p-wave core-neutron scattering 
amplitude in the center-of mass frame:
\begin{eqnarray}
t_1({\bf p}',{\bf p};E)&=&g_1^2 {\bf p}' \cdot {\bf p} D_\pi(E,{\bf 0})
\nonumber\\
&=&\frac{6 \pi}{m_R} \frac{ {\bf p}' \cdot {\bf p}}{1/a_1 - \frac{1}{2} 
r_1 k^2 + i k^3}\,,
\label{eq:t1}
\end{eqnarray}
with $k=\sqrt{2 m_R E}=|{\bf p}'|=|{\bf p}|$ for on-shell scattering.
Consequently we obtain:
\begin{equation}
D_\pi(p)=-\frac{6 \pi}{m_R^2 g_1^2}\frac{1}{r_1 + 3 \gamma_1} 
\frac{1}{p_0 - {\bf p}^2/(2 M_{nc}) + B_1}~\mbox{$+$ regular}\,.
\label{eq:Dpidressed}
\end{equation}
Here $\gamma_1=\sqrt{2 m_R B_1}$ is the solution of
\begin{equation}
\frac{1}{a_1} + \frac{1}{2} r_1 \gamma_1^2 + \gamma_1^3=0\,,
\label{eq:gamma1}
\end{equation}
where $a_1$ is the scattering volume, and $r_1$ the p-wave 
\lq\lq effective range", which, in fact, has dimensions of 1/length. 
Both parameters are required to leading order in the Halo EFT.
The wave-function renormalization for the dressed $\pi$ propagator
can be read off as the residue of the pole in Eq.~(\ref{eq:Dpidressed}):
\begin{equation}
Z_\pi=-\frac{6 \pi}{m_R^2 g_1^2}\frac{1}{r_1 + 3 \gamma_1}\,.
\end{equation}

The propagator (\ref{eq:t1}) has three poles
corresponding to the zeroes of Eq.~(\ref{eq:gamma1}). Using the NLO parameter 
values for the ${}^{10}$Be-n system obtained in Sec.~\ref{sec-numbers} 
below, we find two bound-state poles corresponding to typical momenta 
$\gamma_1 \sim 1/R_{\rm halo}$ and $\gamma_1 \sim 1/R_{\rm core}$. The first is 
that which we identified with the ${}^{11}$Be excited state in the previous 
paragraph. The second is a spurious bound state, which is outside the domain 
of validity of Halo EFT, and is not physically realized in the ${}^{11}$Be 
system. The third solution of Eq.~(\ref{eq:gamma1}) represents a virtual 
state with a typical momentum $\sim 1/R_{\rm halo}$.

The power counting for the propagator $D_\pi(p)$ that we adopt here is that 
of Ref.~\cite{Bd03}. We take $r_1 \sim 1/R_{\rm core}$. The propagator then 
has a pole at $\gamma_1 \sim 1/R_{\rm halo}$, which occurs ``kinematically" 
when $r_1 k^2 \sim 1/a_1$. $a_1$ then must obviously be of order 
$R_{\rm halo}^2 R_{\rm core}$ for such a ``kinematic" pole to occur. We 
note that for $k \sim 1/R_{\rm halo}$ the unitarity piece of the propagator 
has a size $1/R_{\rm halo}^3$. Thus, away from the pole, the dominant 
contribution to the $\pi$ propagator now comes from the bare part (after 
appropriate renormalization) and so:
\begin{equation}
t_1({\bf p}',{\bf p};E)
=\frac{6 \pi}{m_R} \frac{ {\bf p}' \cdot {\bf p}}{-\frac{1}{2} r_1 
(k^2 + \gamma_1^2)}\,,
\label{eq:pwaveundressed}
\end{equation}
where we have also used Eq.~(\ref{eq:gamma1}) to re-express $1/a_1$ 
in terms of $\gamma_1$ and then dropped the $\gamma_1^3$ term relative 
to the (larger) $r_1 \gamma_1^2$ piece.

The result (\ref{eq:pwaveundressed}) can be easily 
re-expressed as:
\begin{equation}
t_1({\bf p}',{\bf p};E)
=-\frac{6 \pi}{m_R^2 r_1} \frac{ {\bf p}' \cdot {\bf p}}{E + B_1}\,.
\label{eq:simplified}
\end{equation}
The amplitude (\ref{eq:simplified})
has only a pole at $E=-B_1 \equiv -\gamma_1^2/(2 m_R)$---a pole that 
corresponds to the 1/2$^-$ state of ${}^{11}$Be. (This pole actually occurs 
on both sheets of the complex $E$-plane, since (\ref{eq:pwaveundressed}) 
exhibits poles in both the lower and upper half of the complex $k$-plane.) 
In contrast, the  spurious deep pole from Eq.~(\ref{eq:t1}) has disappeared 
from the expressions (\ref{eq:pwaveundressed}) and (\ref{eq:simplified}).
Our power counting therefore reproduces the 
spectrum of the $^{11}$Be system.
The p-state wave-function (\ref{eq:LOwfs}) is then obtained, with 
\begin{equation}
A_1=\sqrt{\frac{2 \gamma_1^2}{-r_1}}.
\label{eq:A1}
\end{equation}

The p-wave phase shifts in this theory are given by:
\begin{equation}
k^3 \cot \delta_1=\gamma_1^3 + \frac{1}{2} r_1 (k^2 + \gamma_1^2)\,,
\label{eq:kcotdelta1}
\end{equation}
if no expansions are made. However, for $r_1 \sim 1/R_{\rm core}$ we can, 
once again, drop $\gamma_1^3$ to leading order in a power counting in 
$R_{\rm core}/R_{\rm halo}$, with the result that 
\begin{equation}
\cot \delta_1=\frac{r_1}{2} \left(\frac{1}{k} + \frac{\gamma_1^2}{k^3}\right) 
+ O\left(\frac{R_{\rm core}}{R_{\rm halo}}\right)\,.
\end{equation}
Since $r_1 \gg k, \gamma_1$ we have $\cot \delta_1$ large, which implies 
that $\delta_1$ is approximately zero. Indeed
\begin{equation}
\delta_1=\frac{2}{r_1} \frac{k^3}{k^2 + \gamma_1^2} + O\left(
\frac{R_{\rm core}}{R_{\rm halo}}\right)\,,
\label{eq:delta1pert}
\end{equation}
for all $k \sim \gamma_1 \sim 1/R_{\rm halo}$. Small phase shifts imply 
small unitarity corrections, which is why the imaginary part of $t_1^{-1}$ 
can be treated perturbatively in this regime. 

The only exception to this occurs if we consider $|E-B_1| \sim 
\frac{\gamma_1 B_1}{r_1}$. In that case we are close to the pole and the 
two terms $\sim (R_{\rm core} R_{\rm halo}^2)^{-1}$  in 
Eq.~(\ref{eq:pwaveundressed}) cancel, or come close to doing so. It then 
becomes necessary to  resum the pieces $\sim k^3$ and $\sim \gamma_1^3$ 
which were dropped in order to obtain 
Eq.~(\ref{eq:simplified})~\cite{PP03,Bd03}. In particular, if $B_1 < 0$ 
(i.e. the pole is at positive energy) then these corrections shift the 
pole off the real axis and mean that it represents a resonance. The 
propagator (\ref{eq:t1}) thus describes a p-wave resonance if the pole 
is at positive energy and the width of the resonance at energy $E_R$ is 
of order $k_R/r_1 E_R$~\cite{Be02,Bd03,Sak}. This is thus a narrow resonance 
if $r_1 \sim 1/R_{\rm core}$. The case of p-wave resonances will not be 
discussed further here, since we will restrict ourselves to $B_1 > 0$, as 
is relevant for ${}^{11}$Be. This is the case of a shallow p-wave bound state. 

\subsection{Rescaled fields and naive dimensional analysis}

At this point it is useful to rescale the piece of the Lagrangian which 
encodes the enhanced s- and p-wave interactions. We rewrite ${\cal L}$ in 
terms of fields with non-canonical dimensions, which absorb factors of $g_0$, 
$g_1$, $m$, and $M$~\cite{BS01}. We define:
\begin{equation}
\tilde{\sigma}_s=\sigma_s g_0 m_R; \quad \tilde{\pi}_s=\pi_s g_1 m_R\,.
\label{eq:tildefields}
\end{equation}
The scaling of the absorbed factors $g_0 m_R$ and $g_1 m_R$ can be
obtained by recalling the matching between (\ref{eq:t0sigma}) and 
(\ref{eq:t0ERE}) for s-waves, and in Eq.~(\ref{eq:t1}) for p-waves.
At leading order, this yields:
\begin{equation}
g_0^2 m_R^2 \simeq-\frac{2 \pi \eta_0}{r_0}\,, \quad 
g_1^2 m_R^2 \simeq -\frac{6 \pi\eta_1}{r_1}\,.
\end{equation}
In our counting, we have $r_0 \sim 1/r_1 \sim R_{\rm core}$, which then 
determines how $g_0$ and $g_1$ scale with $R_{\rm core}$. Note also that, 
since $r_0 > 0$ and $r_1 < 0$ for n-${}^{10}$Be interactions, 
we have $\eta_0=-1$, $\eta_1=1$ in this system.

We analyze $M_{nc} {\cal L}$ because this operator is dimension 5, 
and has no powers of the mass in it anymore~\cite{LM97}. 
The expression for the product of the total mass and the pertinent piece 
of the Lagrangian then becomes, in terms of these fields:
\begin{eqnarray}
M_{nc} {\cal L}&=&\frac{1}{g_0^2 m_R^2} \tilde{\sigma}_s^\dagger 
\left[\eta_0 \left(i M_{nc} \partial_t + \nabla^2\right) + M_{nc} 
\Delta_0\right] \tilde{\sigma}_s \nonumber 
 -  \frac{M_{nc}}{m_R} \left[c^\dagger n_s^\dagger \tilde{\sigma}_s 
+ \tilde{\sigma}_s^\dagger n_s c\right]\nonumber\\
&&+\frac{1}{g_1^2 m_R^2}  \tilde{\pi}_s^\dagger  \left[\eta_1
\left(i M_{nc} \partial_t + \nabla^2\right) +  M_{nc} \Delta_1\right] 
\tilde{\pi}_s\nonumber\\ 
&&
- \frac{M_{nc}}{2m_R}\left(\tilde{\pi}_s^\dagger \left[n 
(i\stackrel{\leftrightarrow}{\nabla}) c\right]_{\frac{1}{2},s}
+ \left[c^\dagger (i\stackrel{\leftrightarrow}{\nabla}) n^\dagger
\right]_{\frac{1}{2},s} \tilde{\pi}_s\right)\,,
\end{eqnarray}
where the pieces of ${\cal L}$ that restore Galilean invariance 
have been suppressed.
In terms of these new fields all the coefficients---even those in the 
\lq\lq enhanced" interactions which generate shallow bound states 
(and resonances) are natural. The shallowness of these states is now 
encoded in the fact that the fields associated with them have non-canonical 
dimensions: $[\tilde{\sigma}]=2$, $[\tilde{\pi}]=1$, and non-canonical 
wave-function normalization---even at tree level.

\subsection{Lagrangian: electromagnetic sector}

Photons are then included in the Lagrangian via minimal substitution:
\begin{equation}
\partial_\mu \rightarrow D_\mu=\partial_\mu + i e \hat{Q} A_\mu.
\label{eq:minimal}
\end{equation}
The charge operator $\hat{Q}$ takes different values, depending on whether 
it is acting on a $c$ field or an $n$ field.
$\hat{Q}\, n=0$ for the neutron, and below we denote the eigenvalue of the 
operator $\hat{Q}$ for the $c$ field as 
$\hat{Q}\, c=Q_c\, c$. $Q_c=4$ in the case of interest here, where the core 
is Beryllium-10. 

Here our focus is on electric properties (and form factors), and the 
dominant pieces of the electric response can be derived by looking at how 
the Lagrangian (\ref{eq:HEFT}) is affected by the substitution 
(\ref{eq:minimal}). But, at higher orders in the computation of these 
properties, operators involving the electric field ${\bf E}$ and the fields 
$c$, $n$, $\sigma$, and $\pi$ which are gauge invariant by themselves 
contribute to observables. 
Possible one- and two-derivative operators with one power of the photon 
field are:
\begin{eqnarray}
{\cal L}_{EM}&=&-L_{C0}^{(\sigma)} \sigma_l^\dagger (\nabla^2 A_0 - 
\partial_0 (\nabla \cdot {\bf A}))\sigma_l
- L_{E1}^{(1/2)} \sum_{l l' j} \sigma_l \pi_{l'}^\dagger \left(\left.
\frac{1}{2} l \frac{1}{2} l' \right|1 j\right) (\nabla_j A_0 - \partial_0 A_j)
\nonumber\\
&&
- L_{C0}^{(\pi)} \pi_l^\dagger (\nabla^2 A_0 - \partial_0 (\nabla \cdot 
{\bf A})) \pi_l \nonumber \\
&&
- L_{E1}^{(3/2)} \sum_{ll'j} \sigma_{l} [n (i 
\stackrel{\leftrightarrow}{\nabla}) c]^\dagger_{3/2\, l'} \left(\left.
\frac{1}{2} l \frac{3}{2} l' \right|1 j\right) (\nabla_j A_0 - 
\partial_0 A_j)\,.
\label{eq:nonminimalEM}
\end{eqnarray}
Note that if magnetic properties are to be considered we would also include 
operators involving $\partial_i A_j - \partial_j A_i$ and the neutron, core, 
and bound-state fields. 

The electric interactions in Eq.~(\ref{eq:nonminimalEM}) are gauge invariant 
by themselves, and so we must determine the order at which they occur. 
To do this we rewrite the Lagrangian (\ref{eq:nonminimalEM}) in terms of the 
rescaled fields (\ref{eq:tildefields}). In terms of these fields we assume 
scaling by naive dimensional analysis with respect to the scale 
$R_{\rm core}$ of the operators that appear in $M_{nc} {\cal L}$. We then 
obtain the following scaling of the coefficients written above:
\begin{eqnarray}
L_{C0}^{(\sigma)} &\sim& R_{\rm core}^3 l_{C0}^{(\sigma)} g_0^2 m_R^2
\,,\label{eq:LC0sigma}\\
L_{E1}^{(1/2)} &\sim& R_{\rm core} l_{E1}^{(1/2)} g_0 g_1 m_R^2\,,\\
L_{C0}^{(\pi)} &\sim& R_{\rm core} l_{C0}^{(\pi)}g_1^2 m_R^2\,,\\
L_{E1}^{(3/2)} & \sim & R_{\rm core}^4 l_{E1}^{(3/2)} g_0 m_R\,,
\end{eqnarray}
where the parameters $l_{...}^{...}$ are all of order one.

Below we show that the leading effects in the E1($1/2^+ \rightarrow 1/2^-$) 
matrix element have parametric dependence $R_{\rm halo} 
\sqrt{\frac{R_{\rm core}}{R_{\rm halo}}}$. Including the proper 
wave-function renormalization factors, the operator $\sim L_{E1}^{(1/2)}$ 
yields an effect $\sim R_{\rm core} \sqrt{\frac{R_{\rm core}}{R_{\rm halo}}}$,
and so occurs already at NLO in that quantity. 
Similarly the leading effects 
in the charge-radius-squared of the $1/2^-$ state in ${}^{11}$Be are 
$\sim R_{\rm halo}/r_1 \sim R_{\rm halo} R_{\rm core}$. The operator 
proportional to $L_{C0}^{(\pi)}$ above produces effects in this charge radius of order 
$R_{\rm core}/r_1 \sim R_{\rm core}^2$, and so affects the prediction for 
the p-wave radius at next-to-leading order.
Thus if we desire NLO accuracy for quantities involving the shallow p-wave 
bound state there are two parameters in the Halo EFT description of the 
electric structure of the Beryllium-11 which cannot be fixed from 
${}^{10}$Be-neutron scattering information alone. 

\subsection{Fixing parameters}

\label{sec-numbers}

Using the values $B_0=500$ keV, $B_1=180$ keV from Ref.~\cite{AjK90}, we 
infer $\gamma_0=0.15$ fm$^{-1}$, and $\gamma_1=0.09$ fm$^{-1}$, which are both 
of the expected size $1/R_{\rm halo}$. 
From the power counting discussed in Sec.~\ref{sec:p-wave-int}, we have
at leading order:
\begin{equation}
r_1=-\frac{2}{\gamma_1^2 a_1}.
\label{eq:gamma1lo}
\end{equation}
It follows that if we adopt the value extracted in Ref.~\cite{TB04} from 
experimental data, Eq.~(\ref{eq:pwavescattlength}), we have 
$r_1=(-0.54\pm 0.08)$ fm$^{-1}$. This number should, however, be taken as 
indicative rather than definitive, since in the end we will fit $r_1$ to data 
at both LO and NLO, and deduce corresponding values for $a_1$ from our 
results. But already we see that 
the order of magnitude estimate implied by our counting 
$r_1 \sim 1/R_{\rm core}$ is borne out by the numbers. 

At NLO, Eq.~(\ref{eq:gamma1lo}) is corrected to:
\begin{equation}
r_1=-\frac{2}{\gamma_1^2 a_1} - 2 \gamma_1,
\end{equation}
which, if we again use Eq.~(\ref{eq:pwavescattlength}) to get an idea of 
the effect, alters $r_1$ to $(-0.72\pm 0.08)$ fm$^{-1}$. Such a $\sim$ 30\% 
correction is in line with the anticipated expansion parameter of Halo EFT 
in the ${}^{11}$Be system.

In the s-waves the situation is more straightforward: there we count 
$a_0 \sim R_{\rm halo}$, and $r_0 \sim R_{\rm core}$. In consequence we can set 
$r_0=0$ at LO, and obtain from Eq.~(\ref{eq:gamma0}) 
\begin{equation}
\gamma_0=\frac{1}{a_0}.
\end{equation}
At leading order, in s-waves we have:
\begin{equation}
Z_\sigma=\frac{2\pi\gamma_0}{m_R^2 g_0^2}
\end{equation}
and all other pertinent results can be obtained by taking the 
$r_0 \rightarrow 0$ limit of the formulae in Sec.~\ref{sec-swaves}.

Thus, the parameters in Halo EFT for Beryllium-11 bound states at LO are 
$r_1$, $\gamma_0$ (or equivalently $a_0$), and $\gamma_1$. At NLO these are 
to be supplemented by $r_0$, and the electromagnetic contact interactions for 
the $1/2^+ \rightarrow 1/2^-$ E1 transition and $1/2^-$-state radius.

\section{Results for bound-state observables}
\label{sec:bound}

Using Eq.~(\ref{eq:HEFT}) plus minimal substitution (\ref{eq:minimal}), 
we obtain a Lagrangian that describes interactions amongst the core, the 
neutron, the ground and excited states of the ${}^{11}$Be nucleus, and photons.
In this section we use this Lagrangian to compute the form factor of the 
$1/2^+$ and $1/2^-$ states and the E1 transition from the $1/2^+$ to the 
$1/2^-$ state.

\subsection{s-wave form factor}

\begin{figure}[!t]
\centerline{\includegraphics[width=0.25\columnwidth]{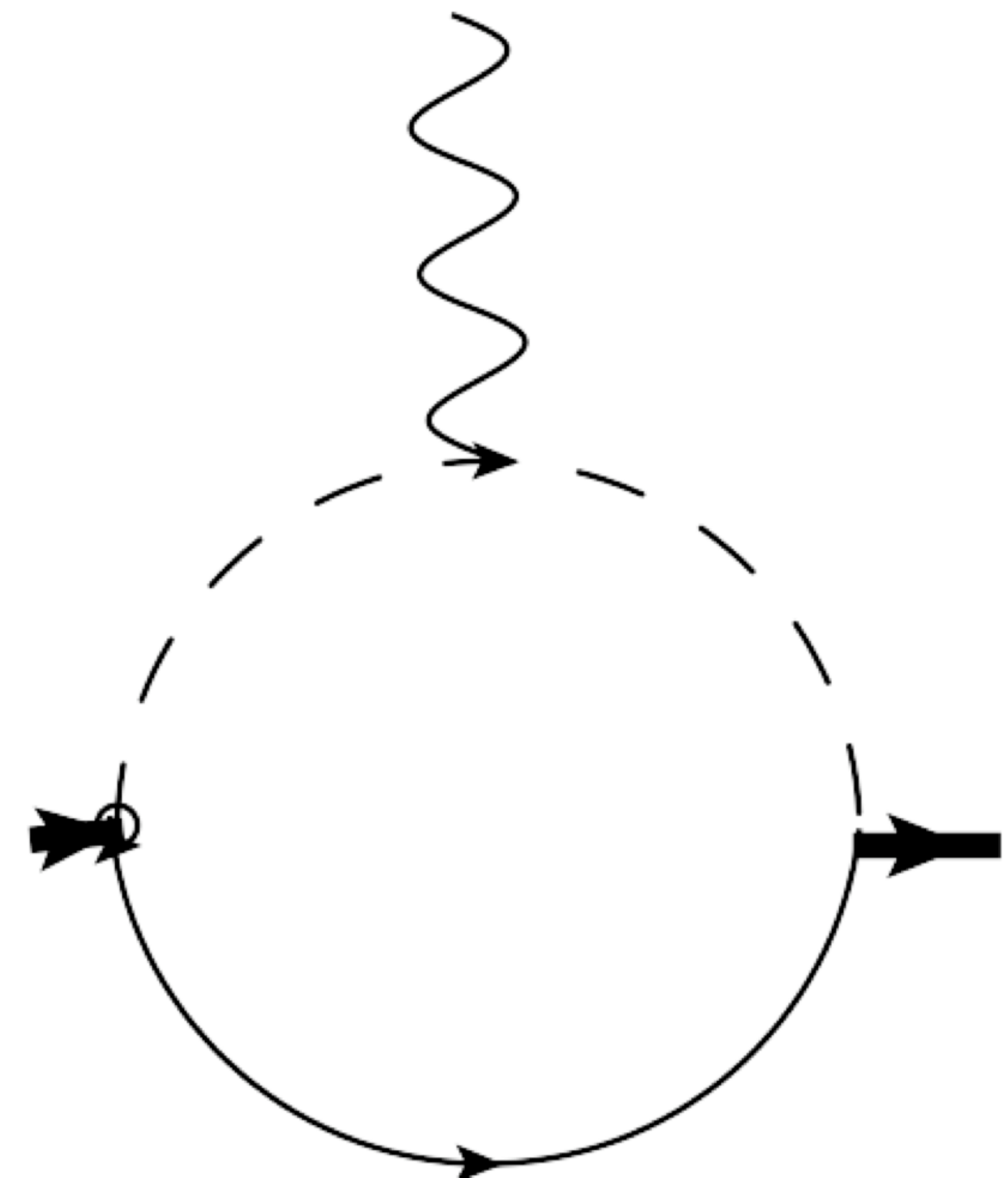}}
\caption{The LO contribution to the irreducible vertex for an $A_0$ photon 
to couple to the field representing the ${}^{10}$Be-neutron s-wave bound 
state. Note that there is no diagram for the photon to couple to the neutron 
as this order, since $Q_n=0$.}
\label{fig:formfactor}       
\end{figure}

The s-wave form factor is computed by calculating the contribution to the 
irreducible vertex for $A_0 \sigma \sigma$ interactions shown in 
Fig.~\ref{fig:formfactor}. This is the only diagram it is necessary to 
consider at leading order. After the application of wave-function 
renormalization, the irreducible vertex for the $A_0$ photon coupling to 
the $\sigma$ state is equal to $-i e Q_c G_E(|{\bf q}|)$, where ${\bf q}$ 
is the three-momentum of the virtual photon. (Such an interpretation is valid 
provided the computation is carried out in the Breit frame, where the 
four-momentum of the virtual photon $q=(0,{\bf q})$.) A straightforward 
calculation yields: 
\begin{equation}
G_E^{(\sigma)}(|{\bf q}|)= \frac{2 \gamma_0}{f|{\bf q}|} \arctan
\left(\frac{f|{\bf q}|}{2 \gamma_0}\right)\,,
\label{eq:Gc}
\end{equation}
with $f=m/M_{nc}=m_R/M$. Note that $G_E(0)=1$, as it should. For the deuteron, 
we have $f=1/2$, and this reduces to the LO result of Ref.~\cite{Ch99}. 

The electric radius of the s-wave state can be extracted according to:
\begin{equation}
G_E^{(\sigma)}(|{\bf q}|) \equiv 1 - \frac{1}{6}\langle r_E^2 
\rangle^{(\sigma)}{\bf q}^2 + \ldots\,,
\label{def:rcqs}
\end{equation}
and an expansion of Eq.~(\ref{eq:Gc}) in powers of $|{\bf q}|$ then yields 
\begin{equation}
\langle r_E^2 \rangle^{(\sigma)}=\frac{f^2}{2 \gamma_0^2}\,.
\label{rcqs}
\end{equation}
Eq.~(\ref{rcqs}) gives the electric radius of the $^{11}$Be ground state 
relative to the electric radius of
$^{10}$Be. 
In order to compare with the experimental radius, we therefore have 
to add our result and the charge radius of $^{10}$Be in quadrature:
\beq
\langle r_E^2 \rangle_{^{11}{\rm Be}}=\langle r_E^2 \rangle_{^{10}{\rm Be}}
+\frac{f^2}{2 \gamma_0^2}\,.
\label{rcqs2}
\eeq
This relation can be derived by writing
the charge distribution of $^{11}$Be as a convolution of the charge 
distribution of $^{10}$Be with that of the $^{10}$Be-n halo system. Using the 
convolution theorem for the Fourier transform, one finds that the total 
rms radius squared can be written as the sum of the squared radii for 
$^{10}$Be and for the $^{10}$Be-n halo system. The latter effect can be
calculated in the Halo EFT.
Inserting the value $\gamma_0=0.15$ fm$^{-1}$ obtained in the previous section
and using the experimental result for the $^{10}$Be charge radius \cite{No09},
we find $\langle r_E^2 \rangle^{1/2}_{^{11}{\rm Be}}=2.40$ fm from this 
leading-order HEFT computation. 
This is 2--3\% smaller than the atomic physics measurement which yields $\langle r_E^2 \rangle^{1/2}_{^{11}{\rm Be}}=2.463(16)$ fm \cite{No09}. 
In fact, comparing our result for $\langle r_c^2 \rangle_{^{11}{\rm Be}}-
\langle r_c^2 \rangle_{^{10}{\rm Be}}$ ($0.19$ fm$^2$) with the experimental 
result 
for this quantity ($0.51(17)$ fm$^2$), the agreement looks poor. 
But, this difference is actually consistent with the nominal $40\%$ 
size of NLO effects when Halo EFT is applied to this system.

At NLO a careful treatment of current conservation, which includes an 
operator associated with gauging the term $\sim \sigma^\dagger \partial_0 
\sigma$ in Eq.~(\ref{eq:HEFT}), still yields $G_E(0)=1$, but also produces 
an increased charge radius, as long as $r_0 > 0$, cf. Ref.~\cite{BS01,Ph02}:
\begin{equation}
\langle r_E^2 \rangle_{^{11}{\rm Be}}=\langle r_E^2 \rangle_{^{10}{\rm Be}}
+\frac{f^2}{2 (1- r_0 \gamma_0) \gamma_0^2}\,.
\end{equation}
Therefore NLO corrections improve the agreement with experiment.
The precise size of the increase is fixed once the s-wave effective range 
$r_0$ is determined, as we shall do in Sec.~\ref{sec-Couldisresults} below. 

We can also obtain from our leading-order calculation a number for the 
mean-square of the relative core-neutron co-ordinate,
$\langle r^2 \rangle$, i.e.:
\begin{equation}
\langle r^2 \rangle=\frac{1}{2 \gamma_0^2}.
\end{equation}
To convert this to a neutron radius, we must insert the conversion 
factor $m_c/(m_c + m_n)$. When this is done we find a LO neutron radius 
for the ${}^{11}$Be ground state of: 
\begin{equation}
\langle r_n^2 \rangle^{1/2}=4.3~{\rm fm}.
\end{equation}
The neutron radius can be measured with a probe that couples only to 
the neutrons. To a very good approximation, the weak gauge boson $Z^0$ 
constitutes such a probe.  Thus one could, in principle, 
measure the rms neutron radius
using parity-violating electron scattering---c.f. the PREX experiment 
for the case of Lead-208~\cite{PREX}.
However, a measurement of parity-violating electron scattering from ${}^{11}$Be
is certainly beyond present-day experimental capabilities.

\subsection{p-wave form factor}

In this subsection, we calculate the charge form factor of the $1/2^-$ 
excited state
in $^{11}$Be. NLO corrections might be expected to 
be smaller there since its binding energy, and so its typical momentum, is 
lower. However, as we shall see, a
counterterm enters already at NLO in this observable. 

\begin{figure}[!t]
\centerline{\includegraphics[width=0.5\columnwidth]{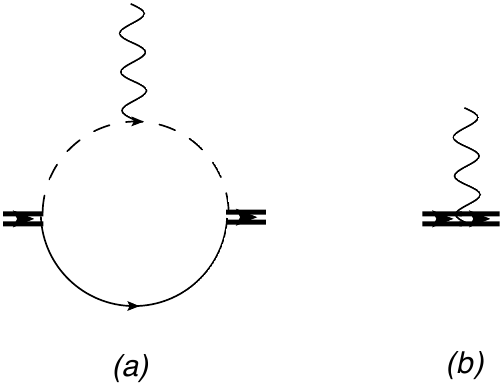}}
\caption{The LO diagrams contributing to the irreducible vertex for 
an $A_0$ photon to couple to the $\pi$ field representing 
the ${}^{10}$Be-neutron p-wave bound state. Diagram (a) is analog 
to the s-wave form factor while diagram (b) arises since the effective 
range $r_1$ enters at leading order for the p-wave state.}
\label{fig:formfactorp}       
\end{figure}

The p-wave form factor is computed by calculating the contribution to the 
irreducible vertex for $A_0 \pi \pi$ 
interactions shown in Fig.~\ref{fig:formfactorp}. There are two diagrams at 
LO. The first diagram is 
analogous to that for the s-wave state while the second diagram represents a direct 
coupling of the photon
to the $\pi$ field. The latter contribution is leading order for the p-wave 
state since the 
effective range $r_1$ is leading order for this state.

As with the self-energy of the $\pi$-field, it is easier to compute the 
irreducible bubble for an $A_0$ photon coupling to the p-wave field 
$n_\beta (i \stackrel{\leftrightarrow}{\nabla}_j) c$. In this way, 
we find that, 
after the application of wave-function renormalization, the irreducible 
vertex for the $A_0$ photon 
coupling to the $1/2^-$ state in the Breit frame can be written as:
\begin{eqnarray}
\langle \pi_s'({\bf p'})| J^0 | \pi_s({\bf p})\rangle &=&-i e 
Q_c \sum_{\alpha \beta i j} \left(\frac{1}{2} \alpha 1 i\right|\left.
\left(\frac{1}{2} 1\right) \frac{1}{2} s\right)
\left(\frac{1}{2} \beta 1 j\right|\left.\left(\frac{1}{2} 1\right) 
\frac{1}{2} s\right) \nonumber\\
&&\qquad\delta_{\alpha \beta}
\left[ G_E^{(\pi)}(|{\bf q}|) \delta_{ij} +\frac{1}{2M_{nc}^2} 
G_Q^{(\pi)}(|{\bf q}|) \left(q_i q_j -
\frac{{\bf q}^2 \delta_{ij}}{3} \right)\right],
\label{def:p-waveFF}
\end{eqnarray}
where ${\bf q}={\bf p'}-{\bf p}$ is the three-momentum of the virtual photon. 
Here we have expressed the form factor in terms of the charge and quadrupole 
form factors of a vector field. Choosing ${\bf q}=|{\bf q}| \hat{z}$, and 
exploiting the fact that the neutron spin is unaffected by the charge 
operators that can occur up to the order we consider here, a brief 
calculation shows 
\begin{equation}
\langle \pi_s'({\bf p'})| J^0 | \pi_s({\bf p})\rangle =-i e Q_c 
\delta_{s's} G_E^{(\pi)}(|{\bf q}|).
\end{equation}
The quadrupole form factor is thus unobservable in the $1/2^-$ state.
It could be observed in a $3/2^-$ state. The lowest
$3/2^-$ state in $^{11}$Be is 2.69 MeV above the ground state. 
However, this state is a n-$^{10}$Be scattering resonance that corresponds to 
typical momenta of order $1/R_{\rm core}$ and thus is outside the range of 
applicability of the Halo EFT. Since the situation could be different
in other one-neutron halo nuclei, we quote the result for the quadrupole 
form factor for completeness:
\begin{equation}
G_Q^{(\pi)}(|{\bf q}|)= \frac{2 M_{nc}^2}{r_1+3\gamma_1}
\frac{3}{4 |{\bf q}|^3 f}
\left(2 |{\bf q}| f \gamma_1 +(|{\bf q}|^2 f^2-4 \gamma_1^2) 
\arctan\left(\frac{f|{\bf q}|}{2 \gamma_1 }\right)\right)\,.
\label{eq:Gq-p}
\end{equation}

In the case of $^{11}$Be, only the charge form factor is observable.
A straightforward calculation yields: 
\begin{equation}
G_E^{(\pi)}(|{\bf q}|)= \frac{1}{r_1+3\gamma_1}\left[r_1 +\frac{1}{|{\bf q}| f}
\left(
2 |{\bf q}| f \gamma_1 +(|{\bf q}|^2 f^2+2 \gamma_1^2) \arctan
\left(\frac{f|{\bf q}|}{2 \gamma_1}\right)\right)\right]\,,
\label{eq:Gc-p}
\end{equation}
where again $f=m/M_{nc}=m_R/M$. For a strict LO result
$r_1+3\gamma_1$ should be replaced by $r_1$ in these expressions.

We have $G_E^{(\pi)}(0)=1$, as required by charge conservation.
The electric radius of the p-wave state relative to the $^{10}$Be ground state
can be extracted according to Eq.~(\ref{def:rcqs}), and we obtain the electric 
radius for the case $r_1 \sim \gamma_1$
from an expansion of Eq.~(\ref{eq:Gc-p}) in powers of $|{\bf q}|$:
\begin{equation}
\langle r_E^2 \rangle^{(\pi)}=-\frac{5f^2}{2 \gamma_1 (3\gamma_1+r_1)}\,.
\label{rcqs-p}
\end{equation}
At LO in the situation of interest here $|r_1| \gg \gamma_1$, we should 
replace this by:
\begin{equation}
\langle r_E^2 \rangle^{(\pi)}=-\frac{5 f^2}{2 \gamma_1 r_1}=
\frac{5 f^2 a_1^{\rm LO} \gamma_1}{4}\,.
\label{eq:LOpradresult}
\end{equation}
This is, as promised above, $\sim R_{\rm halo} R_{\rm core}$.

It is interesting to examine why the p-wave radius is more sensitive to 
short-distance physics than the s-wave one. 
From the co-ordinate space point of view we find that at LO the radius of the 
p-wave state must be calculated as:
\begin{equation}
\langle r_E^2 \rangle^{(\pi)}=A_1^2 f^2 \int_0^\infty \, dr \,r^2 
\left(1 + \frac{1}{\gamma_1 r}\right)^2 e^{-2 \gamma_1 r}\,.
\end{equation}
Inserting $A_1$ from Eq.~(\ref{eq:A1}) yields Eq.~(\ref{eq:LOpradresult}). 
Since the integral is finite, we can compute the contribution to it from 
values of $r \ll 1/\gamma_1$:
\begin{equation}
\langle r_E^2 \rangle^{(\pi)}_{SD}=A_1^2 f^2 \int_0^{R_{\rm core}} \, dr \, 
r^2 \left(1 + \frac{1}{\gamma_1 r}\right)^2 e^{-2 \gamma_1 r}\,,
\label{eq:rE2SD}
\end{equation}
which produces, for $R_{\rm core} \ll 1/\gamma_1$, a short-distance
fraction of the total result of:
\begin{equation}
\frac{\langle r_E^2 \rangle^{(\pi)}_{SD}}{\langle r_E^2 \rangle^{(\pi)}} 
\sim R_{\rm core} \gamma_1\,.
\end{equation}
The parametric dependence of the short-distance contribution on $R_{\rm core}$ 
is in accord with the result obtained in the previous section for the 
corresponding counterterm using naive dimensional analysis in our rescaled 
Lagrangian. 

It might seem counterintuitive that there is a short-distance contribution 
to $\langle r_E^2 \rangle^{(\pi)}$ already at NLO---especially when the 
corresponding effect does not occur in $\langle r_E^2 \rangle^{(\sigma)}$ 
until N$^3$LO (see Eq.~(\ref{eq:LC0sigma}) and Ref.~\cite{Ch99}). The physics 
of this is, however, quite straightforward. It is associated with the 
propensity of the p-wave state's probability distribution to be drawn in 
to shorter distances than the s-wave one, as it gets caught between the 
attractive potential that produces the excited state of ${}^{11}$Be and the 
centrifugal barrier. Observables associated with a shallow p-wave bound state 
will, therefore, generically exhibit counterterms at lower order than those 
of their s-wave counterparts. 

Numerical evaluation of the LO expression (\ref{eq:LOpradresult})  leads to 
the prediction for the charge radius of the 
$^{11}$Be p-wave state relative to the $^{10}$Be ground state
$\langle r_c^2 \rangle^{(\pi)}=0.36$ fm$^{2}$ at LO. (Here we have used as input the value $r_1=-0.66$ fm, see the next section.)
A trivial way to estimate 
the size of NLO corrections is to use the result for $A_1$ in the limit 
$|r_1| \sim \gamma_1$:
\begin{equation}
A_1=\sqrt{-\frac{2 \gamma_1^2}{r_1 + 3 \gamma_1}}
\label{eq:A1difft}
\end{equation}
instead of Eq.~(\ref{eq:A1}). This yields a 20\%
correction at NLO in agreement with the expectation from the power counting, 
although it must be remembered that the entire NLO result in the situation 
where $|r_1| \gg \gamma_1$ includes the contributions of the operator 
$\sim \pi_s^\dagger \nabla^2 A_0 \pi_s$. Thus the only prediction we can 
offer here is a leading-order one. Using again the experimental result for the 
$^{10}$Be charge radius \cite{No09}, we
predict the electric radius of the $1/2^-$ state as:
\begin{equation}
\langle r_E^2 \rangle^{1/2}_{^{11}{\rm Be}^*}=(2.43 \pm 0.1)\, {\rm fm}
\label{eq:rE2pi}
\end{equation}
where our error bar comes from the above estimate of NLO effects. To our knowledge there is, as yet, no experimental determination of the 
charge radius of this state. 
Note that in Eq.~(\ref{eq:rE2pi}) we are assuming that the short-distance 
effects in $\langle r_E^2 \rangle^{1/2}_{^{11}{\rm Be}^*}$ scale with $f$, as 
suggested by the renormalization-group argument summarized in 
Eq.~(\ref{eq:rE2SD}). Short-distance effects which do not involve a factor 
of $f$, e.g., modification of the proton distribution of ${}^{10}$Be in 
the ${}^{11}$Be excited state due to a non-recoil effect, could have an impact 
on our final result that is larger than $0.1$ fm.

\subsection{E1 transition: $1/2^+ \rightarrow 1/2^-$ state}

\begin{figure}[!h]
\centerline{\includegraphics[width=0.6\columnwidth]{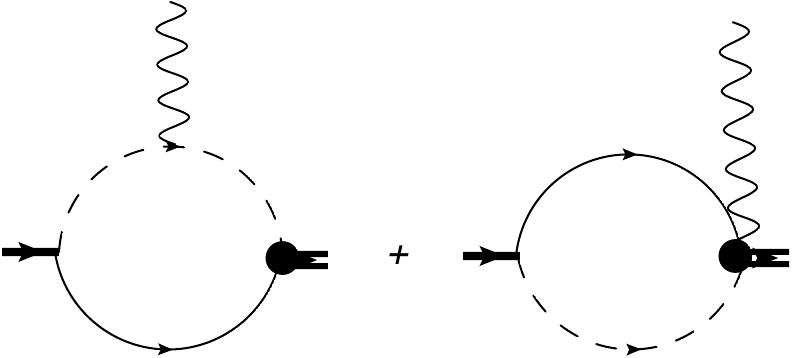}}
\caption{The two diagrams needed for the irreducible vertex that governs the s-to-p-state transition, 
$\Gamma_{j \mu}$ in Halo EFT at leading order.}
\label{fig:Gammajmu}       
\end{figure}

Now we consider the E1 transition from the $1/2^+$ state to the $1/2^-$ state. 
The irreducible vertex for this transition is depicted in 
Fig.~\ref{fig:Gammajmu}. We compute the transition for a photon of arbitrary 
four momentum $k=(\omega,{\bf k})$, and the sum of diagrams yields 
$-i \Gamma_{s' s  \mu}$ where $s'$ ($s$) is the spin projection of the $1/2^-$ 
($1/2^+$) state and $\mu$ is the polarization index of the photon. 

We first observe that both the diagrams depicted in Fig.~\ref{fig:Gammajmu} 
are divergent, but that the divergences cancel, as they should since gauge 
invariance precludes us from writing down any contact interaction that 
contributes to this observable at leading order. Current conservation at 
LO can be explicitly checked, and we find that, as long as both diagrams are 
considered~\cite{HP10}:
\begin{equation}
k^\mu \Gamma_{s's \mu}=0\,.
\label{eq:currentcons}
\end{equation}
Note that if only the long-distance E1 mechanism on the left-hand side of 
Fig.~\ref{fig:Gammajmu} is considered, as was done, for example, in 
Ref.~\cite{TB08}, then current conservation is not satisfied, and it appears 
that some input from short-distance physics is needed in order to define the 
prediction for this observable. 

Since we are considering electric properties, and the spin of the neutron is 
not affected by the photon interaction, it is, once again, convenient to 
re-express $\Gamma_{s's \mu}$ in terms of the vertex function for specific 
components of the p-wave interaction that appears at the $\pi$-neutron-core 
vertex:
\begin{equation}
\Gamma_{s's\mu}=\sum_j \left(\frac{1}{2} s 1 j\right|\left.\left(\frac{1}{2} 
1\right) \frac{1}{2} s'\right) \tilde{\Gamma}_{j \mu}\,.
\label{eq:clebschforE1}
\end{equation}
We note that if we examine the case $s'=s$ then only $j=0$
contributes to the sum in Eq.~(\ref{eq:clebschforE1}), 
and we have
\begin{equation}
-\Gamma_{++ \mu}=\Gamma_{--\mu} =\frac{1}{\sqrt{3}} \tilde{\Gamma}_{3 \mu}\,,
\end{equation}
where $\pm$ indicates the spin projection $\pm 1/2$.

For real photons we have ${\bf k} \cdot {\bf \epsilon}=0$ and 
we choose ${\bf k} \cdot {\bf p}=0$, with ${\bf p}$ the incoming momentum of 
the s-wave state. With these restrictions we can write the space-space 
components of the $\tilde{\Gamma}$ vertex function as:
\begin{equation}
\tilde{\Gamma}_{ji}=\delta_{ji} \Gamma_E + k_j p_i \Gamma_M\,.
\end{equation}
We can, without loss of generality, choose the photon to be traveling in the 
$\hat{z}$ direction, and 
it then follows that $\tilde{\Gamma}_{33}=\Gamma_E$.

Using the definition of B(E1) strength (see, e.g.~\cite{TB05}) we find that 
the B(E1) strength for this transition is related to the renormalized, 
irreducible vertex $\bar{\Gamma}_E$ by:
\begin{equation}
{\rm B(E1)}=\frac{3}{4 \pi} \left(\frac{\Gamma_{++ 3}}{\omega}\right)^2=
\frac{1}{4 \pi} \left(\frac{\bar{\Gamma}_E}{\omega}\right)^2\,,
\label{eq:BE1formula}
\end{equation}
where 
\begin{eqnarray}
\bar{\Gamma}_E &\equiv& \sqrt{Z_\sigma Z_\pi} \Gamma_E
=\frac{1}{m_R^2 g_0 g_1} \sqrt{\frac{-12 \pi^2 \gamma_0}{r_1}}  \Gamma_E
\end{eqnarray}
at leading order. 

Current conservation (\ref{eq:currentcons}) then provides an alternative 
way to calculate $\Gamma_E$, it tells us that:
\begin{equation}
\omega \Gamma_{j0}={\bf k}_j \Gamma_E\,.
\end{equation}
But, for $\Gamma_{j0}$, the diagram on the right of Fig.~\ref{fig:Gammajmu} 
need not be considered, and so
\begin{equation}
\Gamma_{j0}({\bf k}) \sim  \int d^3r \frac{u_1(r)}{r} Y_{1j}(\hat{r}) 
e^{i {\bf k} \cdot {\bf r}} \frac{u_0(r)}{r}\,,
\label{eq:Gamma0j}
\end{equation}
where $u_1$ and $u_0$ are the leading-order wave functions of the s- and 
p-wave states, given by Eq.~(\ref{eq:LOwfs}).
As $|{\bf k}| \rightarrow 0$ Eq.~(\ref{eq:Gamma0j}) reduces to:
\begin{equation}
\Gamma_{j0}({\bf k}) \sim {\bf k}_j \int dr\,r \, u_1(r)\, u_0(r)\,,
\label{eq:Gammaj0}
\end{equation}
an equation in which, of course, the integral is the canonical form of the 
E1 matrix element. 

We performed both the momentum-space calculation of $\tilde{\Gamma}_{33}=\Gamma_E$, 
and the calculation of $\Gamma_E$ via the co-ordinate space integral 
(\ref{eq:Gammaj0}) (with appropriate factors). The divergences cancel in the 
former calculation, as expected, and the same result is obtained from either integral. Inserting 
the result in Eq.~(\ref{eq:BE1formula}) we find:
\begin{equation}
{\rm B(E1)}=\frac{Z_{eff}^2 e^2}{3 \pi} \frac{\gamma_0}{-r_1} \left[
\frac{2 \gamma_1 + \gamma_0}{(\gamma_0 + \gamma_1)^2}\right]^2
\label{eq:BE1}
\end{equation}
is the leading-order Halo EFT result
with
\beq
Z_{eff}=\frac{m_R}{M} Q_c \equiv f Q_c \approx 0.366
\eeq
the effective charge.

No cutoff parameter is needed in order to get a finite result for B(E1): 
our value is finite without regularization, c.f. Ref.~\cite{TB08}. We note that the result 
(\ref{eq:BE1}) is ``universal" in the sense that it applies to any E1 
s-to-p-wave transition in a one-neutron halo nucleus. Once $r_1$, $\gamma_1$, 
and $\gamma_0$ are known for a given one-neutron halo the prediction 
(\ref{eq:BE1}) is accurate up to corrections of order $R_{\rm core}/R_{\rm halo}$.

We now have to deal with the issue that we do not have a value for $r_1$ 
that has been obtained solely from an EFT calculation and data. So here we 
set $r_1$ in order to reproduce the experimental number  (\ref{eq:CEXBE1}) 
obtained in Ref.~\cite{Su07}. This produces:
\begin{equation}
r_1^{\rm LO}=-0.66~{\rm fm}^{-1},
\label{eq:r1LO}
\end{equation}
where we do not bother to propagate the error bars from the experiment, 
since NLO effects are presumably a much larger source of uncertainty. 
This corresponds to $a_1=374~{\rm fm}^3$. We note, that the value of $r_1$ (\ref{eq:r1LO}) 
lies between the two values extracted when we adopt the value of $a_1$ 
from Ref.~\cite{TB04} as given and obtained $r_1$ using formulae of LO and 
NLO precision in the $R_{\rm core}/R_{\rm halo}$ expansion (see Sec.~\ref{sec-numbers}). 

Short-distance effects enter B(E1) in these NLO corrections. The B(E1) 
($1/2^+ \rightarrow 1/2^-$) transition therefore cannot be predicted at 
NLO: a counterterm appears at that order. This can be seen either from the 
presence of the operator $\sim L^{(1/2)}_{E1}$ in Eq.~(\ref{eq:nonminimalEM}), or from 
a co-ordinate space argument similar to the one made in the previous section 
for the $1/2^-$ state's charge radius. In the case of the E1 transition we 
have the co-ordinate space integral~\cite{TB04,TB05}:
\begin{equation}
R_{01}^{(1)} \equiv A_0 A_1 \int_0^\infty dr\, r\, e^{-\gamma_1 r} \left(1 + 
\frac{1}{\gamma_1 r}\right) e^{-\gamma_0 r}=2\sqrt{-\frac{\gamma_0}{r_1}}
\frac{\gamma_0 + 2 \gamma_1}{(\gamma_0 + \gamma_1)^2}\,.
\end{equation}
Again, it is finite, and this time it has a short-distance contribution 
\begin{equation}
{R_{01|SD}^{(1)}} \sim 2 \sqrt{-\frac{\gamma_0}{r_1}} R_{\rm core}
\end{equation}
which is a fraction $\sim R_{\rm core}/R_{\rm halo}$ of the total. 

Comparing this calculation with a shell-model treatment of ${}^{11}$Be, it 
is clear that one effect which is
subsumed into the NLO counterterm $L^{(1/2)}_{E1}$ is the transition of a 
neutron
from a $d_{5/2}$ to a $p_{3/2}$ orbital, with that neutron coupled to the
$2^+$ state of ${}^{10}$Be. This $2^+$ state is 3.4 MeV
above the ${}^{10}$Be ground state, so the dynamics associated with it takes 
place
at distances $R_{\rm core}$. Hence in our EFT it can only appear in 
short-distance operators
such as that multiplying $L^{(1/2)}_{E1}$ in Eq.~(\ref{eq:nonminimalEM}). 
The computation of Ref.~\cite{Mi83} suggests that such a contribution
reduces the E1 matrix element by $\sim 30$\%, which is the anticipated size
of an NLO effect when the $R_{\rm core}/R_{\rm halo}$ expansion is employed in 
the ${}^{11}$Be system.

There are other effects of a similar size that will affect 
B(E1) at NLO. Specifically, there are 
NLO corrections from the wave-function renormalization factors associated 
with the s-wave and p-wave fields. Both tend to increase B(E1) over the 
leading-order prediction. We choose to adjust $r_1$ to reproduce 
B(E1) ($1/2^+ \rightarrow 1/2^-$) already at leading order, and then rely 
on the NLO counterterm to cancel these NLO corrections. That this is a 
reasonable strategy is a testable hypothesis, since 
the counterterm that appears here at NLO also plays a role in Coulomb 
dissociation of ${}^{11}$Be. It is to that process that we now turn our 
attention. 

\section{Photodisintegration of ${}^{11}$Be into ${}^{10}$Be and  
a neutron}

\noindent

In this section we consider the photodisintegration of the Beryllium-11 
nucleus to a Beryllium-10 nucleus plus a neutron. In practice 
this process is measured using Coulomb excitation of the ${}^{11}$Be nucleus, with the 
two reactions connected within the equivalent-photon approximation. 

Once again, we first compute photodisintegration into the direct product of 
a core state of definite polarization and a neutron of definite spin. The 
separation into states of definite total angular momentum will be done 
after that calculation is complete.

\begin{figure}[ht]
\centerline{\includegraphics[width=0.75\columnwidth]{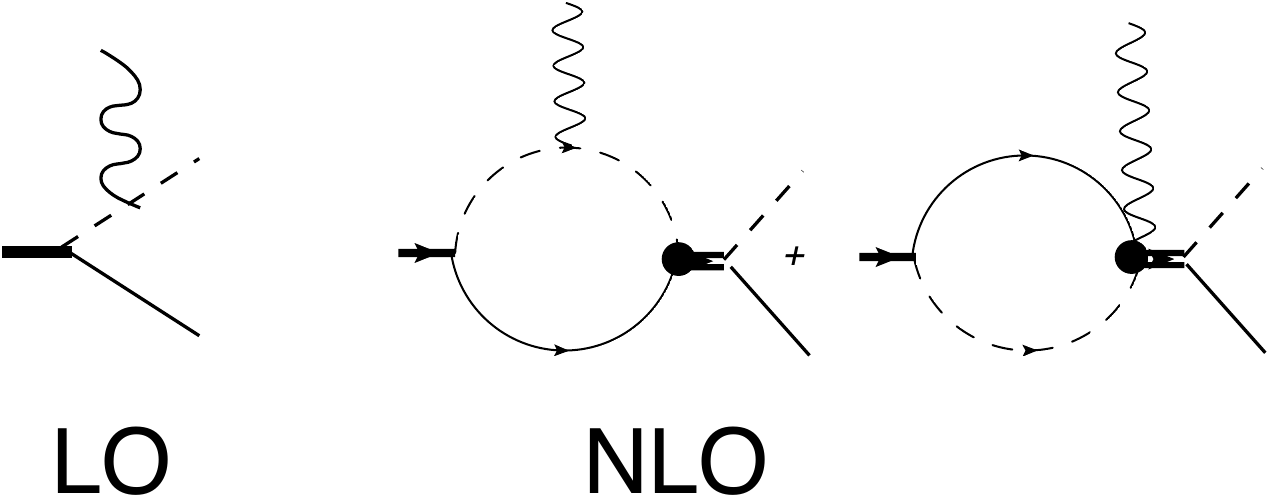}}
\caption{The two diagrams representing photodissociation of the ${}^{11}$Be 
ground (s-wave) state. As we will show below, the first diagram, denoted LO 
here, is dominant over diagrams involving p-wave final-state interactions.}
\label{fig:photodis}       
\end{figure}

There are two contributions to this process, as depicted in 
Fig.~\ref{fig:photodis}. The first diagram, denoted ``LO" in the figure, 
corresponds to the plane-wave impulse approximation contribution. Evaluation 
of the relevant Feynman graph yields, for the zeroth-component of the 
four-current:
\begin{equation}
{\cal M}_0^{(a)}=\frac{e Q_c g_0 2 m_R}{\gamma_0^2 + \left({\bf p}' - 
\frac{m}{M_{nc}} {\bf k}\right)^2}\,,
\label{eq:M0a}
\end{equation}
where ${\bf p}'$ is the relative momentum of the outgoing $nc$ pair. 

The second diagram pair of diagrams, (b), includes final-state interactions 
between the neutron and the core. (In fact, these interactions are only 
present in the $J=1/2$ channel, a point we will have to deal with below.) 
The diagram with final-state interactions can be written as:
\begin{equation}
{\cal M}_0^{(b)}=e Q_c g_0 (2 m_R)^2 \frac{-6 \pi}{m_R} \frac{1}{p'^3 \cot 
\delta_1(p') - i p'^3} p'_j  L_j\,,
\label{eq:M0b}
\end{equation}
with $p'^3 \cot \delta_1(p')=-\frac{1}{a_1} + \frac{1}{2} r_1 p'^2$ if we 
do not yet expand for the case $r_1 \sim 1/R_{\rm core}$. In Eq.~(\ref{eq:M0b})
repeated indices are summed over and $L_j$ is the (vector) integral
\begin{equation}
L_j \equiv \int \frac{d^{d-1}p}{(2 \pi)^{d-1}} \left(p_j + \frac{m}{M_{nc}} 
k_j\right) \frac{1}{p'^2 - \left({\bf p} + \frac{m}{M_{nc}} {\bf k}\right)^2 
+ i \eta}\frac{1}{\gamma_0^2 + {\bf p}^2}\,.
\end{equation}
This can be converted into a co-ordinate space integral, with the result:
\begin{equation}
L_j=\frac{1}{4 \pi} \hat{k}_j \int \, dr \, r^2 \left(i p'-\frac{1}{r}\right) 
\frac{e^{i p' r}}{r}  
j_1 \left(\frac{m}{M_{nc}} \omega r\right) \frac{e^{-\gamma_0 r}}{r}
\,.
\end{equation}
Thus, 
\begin{equation}
{\cal M}_0^{(b)}=-\frac{e Z_{eff} g_0 2 m_R \omega}{p'^2} 
\frac{e^{2 i \delta_1(p')} 
-1}{2 i} \hat{p}' \cdot \hat{k} \int dr (i p' r - 1) e^{i p' r}e^{-\gamma_0 r} 
+ O\left(\frac{m^3 \omega^3}{M_{nc}^3}r^3\right)\,.
\end{equation}

Meanwhile, the piece of ${\cal M}_0^{(a)}$ corresponding to a final-state 
$nc$ p-wave may be rewritten as:
\begin{equation}
{\cal M}_0^{(a)}=e Z_{eff} g_0 2 m_R \omega \hat{k} \cdot \hat{p}' \int dr 
\, r^2 j_1(p'r) e^{-\gamma_0 r}\,.
\end{equation}
In the presence of final-state interactions, it is most useful to write the 
combination ${\cal M}_0^{(a)} + {\cal M}_0^{(b)}$ in the form:
\begin{equation}
{\cal M}_0=e Z_{eff} g_0 2 m_R \omega \hat{k} \cdot \hat{p}' e^{i \delta_1(p')} 
\int dr \,  r^2 \left[j_1(p' r) \cos \delta_1(p') + n_1(p' r) \sin \delta_1(p')
\right] e^{-\gamma_0 r}\,.
\label{eq:M0}
\end{equation}
Note that the sign of the spherical Neumann function used here is opposite to that chosen in many textbooks. We define:
\begin{equation}
n_1(x)=\frac{\cos(x)}{x^2} + \frac{\sin(x)}{x}\,.
\end{equation}

Next we follow Typel and Baur in Refs.~\cite{TB04,TB05} and re-express Eq.~(\ref{eq:M0}) in 
terms of dimensionless integrals:
\begin{equation}
{\cal M}_0=\frac{e Z_{eff} g_0 2 m_R \omega}{\gamma_0^3} \hat{k} \cdot \hat{p}' 
e^{i \delta_1(p')} \left[-f_1(y) \cos(\delta_1(p')) + 
f_2(y)\sin(\delta_1(p'))\right]\,,
\label{eq:M0new}
\end{equation}
where $y=p'/\gamma_0$, and
\begin{eqnarray}
f_1(y) &\equiv& -\int_0^\infty dx \, x^2 e^{-x} j_1(yx)=-\frac{2 y}{(1+y^2)^2}
\,,\\
f_2(y) &\equiv& \int_0^\infty dx \, x^2 e^{-x} n_1(yx)=\frac{1 + 3 y^2}{
(1+y^2)^2 y^2}\,.
\end{eqnarray}

This calculation corresponds to the insertion of a plane-wave for the 
electromagnetic field, followed by an expansion of the effects to first 
order in ${\bf r}$. The resulting matrix element differs from that of 
the dipole operator $|{\bf r}| Y_{10}(\hat{r})$ by a factor of  
$\sqrt{3/(4 \pi)}$ and a $1/\omega$. Thus to get the physical matrix element 
of the E1 strength ${\cal M}_{E1}$ we multiply ${\cal M}_0$ by 
$\sqrt{3/(4 \pi)}$ and divide by $\omega$.
We must also multiply by the field renormalization $\sqrt{Z_\sigma}$, for the 
initial state. This also justifies the factors appearing in 
Eq.~(\ref{eq:BE1formula}).

This produces the E1 coupling to the physical $\sigma$ field of a plane wave, 
in the absence of neutron spin, $\overline{\cal M}_{E1}$. To get the result 
for a final state of good total angular momentum we need to couple 
$\overline{\cal M}_{E1}$ to the neutron spinor, and then form states of good 
total $J$. This leads to two different $\overline{\cal M}_{E1}$ amplitudes, 
one for $J=1/2$ and one for $J=3/2$. 

Only the $J=1/2$ amplitude has final-state interaction effects. Thus there we 
have:
\begin{equation}
\overline{\cal M}_{E1}^{(1/2)}=A_0 \frac{e Z_{eff} \sqrt{3}}{\gamma_0^3} 
e^{i \delta_{(1/2)}(p')}  \frac{2 y^3 \cos(\delta_{(1/2)}(p')) + 
(1 + 3 y^2)\sin(\delta_{(1/2)}(p'))}{y^2(1+y^2)^2} \frac{1}{\sqrt{3}}\,,
\label{eq:ME11/2}
\end{equation}
where the irrelevant spin-dependent phase in the Clebsch-Gordan 
coefficient has been dropped.
Here we have chosen the photon momentum $\hat{k}$ to be in the z-direction. 
The $Y_{10}(\hat{p}')$ that then occurs in ${\cal M}_0$ is coupled to the 
final-state neutron spin to produce a 
${\cal Y}_{(\frac{1}{2} 1)\frac{1}{2} s}(\hat{p}')$, where $s$ is the spin 
projection of the initial state. The coupling of $Y_{10}(\hat{p}')$ to the 
neutron spinor yields the Clebsch-Gordon coefficient 
$(\frac{1}{2} s 1 0|(\frac{1}{2} 1) \frac{1}{2} s)=\sqrt{\frac{1}{3}}$.
 The projection onto the $J=1/2$ continuum $nc$ state then removes this 
${\cal Y}$. Note that physically 
the final-state spin projection of the core-neutron state is equal to that 
of the initial state because the $m_s$ of the neutron is not changed by the 
electric transition. Meanwhile, the angular-momentum projection of the 
nuclear state is unaffected by the photon, since it is taken to be in the 
z-direction. Finally, the final-state strong interactions conserve the 
sum $m_s + m_l$, and so $m_j'$, the angular-momentum projection of the 
final state is equal to $s$, the spin projection of the neutron in the 
initial state.

For the spin-$3/2$ part final-state interactions are natural, and so are 
suppressed by three powers of $R_{\rm core}/R_{\rm halo}$. Thus in this piece 
of the amplitude we neglect diagram (b),  which is equivalent to taking 
$\delta_1 \rightarrow 0$ in Eq.~(\ref{eq:M0new}). Alternatively, we can project 
out the E1, $J=3/2$, piece of the result from Eq.~(\ref{eq:M0a}) for 
${\cal M}_0^{(a)}$, and multiply by $\sqrt{Z_\sigma}$ to find:
\begin{equation}
\overline{\cal M}_{E1}^{(3/2)}=A_0 e Z_{eff} \sqrt{3} \frac{2 p'}{(p'^2 + 
\gamma_0^2)^2} \sqrt{\frac{2}{3}}\,,
\label{eq:ME13/2}
\end{equation}
where an irrelevant spin-dependent phase again has been dropped.
In this case the coupling of $Y_{10}(\hat{p}')$ to the neutron spinor yields 
a Clebsch-Gordan coefficient $(\frac{1}{2} s 1 0|(\frac{1}{2} 1) 
\frac{3}{2} s)=\sqrt{\frac{2}{3}}$.
Finally we convert to the physical observable $\frac{\rm{dB(E1)}}{dE}$, by:
\begin{equation}
{\rm dB(E1)}=\left(|\overline{\cal M}_{E1}^{(1/2)}|^2 + 
|\overline{\cal M}_{E1}^{(3/2)}|^2\right) \frac{d^3p'}{(2 \pi)^3}\,.
\label{eq:diffBE1}
\end{equation}
Inserting Eqs.~(\ref{eq:ME11/2}) and (\ref{eq:ME13/2}) in 
Eq.~(\ref{eq:diffBE1}) produces:
\begin{equation}
\frac{d{\rm B(E1)}}{dE}=e^2 Z_{eff}^2 \frac{m_R}{2 \pi^2} A_0^2 
\left(\frac{[2 p'^3 \cos(\delta_{(1/2)}(p')) + 
(\gamma_0^3 + 3 \gamma_0 p'^2)\sin(\delta_{(1/2)}(p'))]^2}{p'^3 
(p'^2 + \gamma_0^2)^4}  + \frac{8 p'^3}{(p'^2 + \gamma_0^2)^4}\right)\,,
\label{eq:suitforpert}
\end{equation}
where $E=p'^2/(2m_R)$ is the kinetic energy of the neutron and $^{10}$Be
in the center-of-mass frame.
This expression can be rendered more straightforward to work with if we use 
the identity
$\cot^2 \delta+1=\sin^{-2} \delta$ to write things in terms of $p'^3 \cot 
\delta_{(1/2)}(p')$:
\begin{equation}
\frac{d{\rm B(E1)}}{dE}=e^2 Z_{eff}^2 \frac{m_R}{2 \pi^2} A_0^2 
\left(\frac{p'^3 [2 p'^3 \cot(\delta_{(1/2)}(p')) + 
\gamma_0^3 + 3 \gamma_0 p'^2]^2}{[p'^6 + p'^6 \cot^2(\delta_{(1/2)}(p'))] 
(p'^2 + \gamma_0^2)^4}  + \frac{8 p'^3}{(p'^2 + \gamma_0^2)^4}\right)\,,
\label{eq:nonpert}
\end{equation}
which makes it clear that the first term (the contribution of the $J=1/2$ 
channels) does {\it not} diverge as $p' \rightarrow 0$.

The expression (\ref{eq:nonpert}) is true to all orders in final-state 
interactions, and for any value of the s-wave asymptotic normalization $A_0$. 
However, as explained above, in the ${}^{11}$Be system, final-state 
interactions---even in the $J=1/2$ channel---are weak, and can be considered sub-leading. Also,  
$A_0=\sqrt{2 \gamma_0}$ at leading order, but then receives higher-order corrections in the case of a non-zero $r_0$. Thus we now 
provide order-by-order expressions for $d\rm{B(E1)}/dE$ in the 
$R_{\rm core}/R_{\rm halo}$ expansion.

First, at LO we have $\delta_{(1/2)}(p')=0$, and $A_0=\sqrt{2 \gamma_0}$, so:
\begin{equation}
\frac{d{\rm B(E1)}}{dE}^{\rm LO}=e^2 Z_{eff}^2 \frac{3 m_R}{2 \pi^2} \frac{8 \gamma_0 p'^3}{(p'^2 + \gamma_0^2)^4}\,.
\label{eq:dBE1dELO}
\end{equation}

The NLO correction comes from two sources. The first is the shift of $A_0$ 
to larger values due to a $r_0 > 0$, which tends to increase the B(E1) 
strength. Second, final-state interactions make $\delta_{(1/2)}(p') \neq 0$. 
Employing Eq.~(\ref{eq:delta1pert}) for $\delta_{(1/2)}(p')$ in 
Eq.~(\ref{eq:suitforpert}) and keeping only terms up to first order in that 
phase shift, the total NLO result becomes:
\begin{eqnarray}
\frac{d{\rm B(E1)}}{dE}^{\rm NLO}&=&e^2 Z_{eff}^2 \frac{3 m_R}{2 \pi^2} 
\frac{8 \gamma_0 p'^3}{(p'^2 + \gamma_0^2)^4} \left(\frac{A_0^2}{2 \gamma_0}
+ \frac{2 \gamma_0}{3 r_1}\frac{\gamma_0^2  + 3 p'^2}{p'^2 + \gamma_1^2}
\right)\\
&=&e^2 Z_{eff}^2 \frac{3 m_R}{2 \pi^2} \frac{8 \gamma_0 p'^3}{(p'^2 + 
\gamma_0^2)^4} \left(1+r_0 \gamma_0 + 
\frac{2 \gamma_0}{3 r_1}\frac{\gamma_0^2  
+ 3 p'^2}{p'^2 + \gamma_1^2}\right)\,,
\label{eq:dBdE1NLO}
\end{eqnarray}
where in the second line we have used (\ref{eq:A0NLO}) and kept only the term 
of $O(r_0 \gamma_0)$. Accurate measurements of the Coulomb dissociation 
spectrum therefore provide information on the s-wave ${}^{10}$Be effective 
range, if the p-wave effective range is already fixed from another observable.

This sort of expansion of Eq.~(\ref{eq:nonpert}) can be carried to arbitrarily 
high orders in $R_{\rm core}/R_{\rm halo}$. However, it must be remembered that 
the counterterm that affects ${\rm B(E1)} (1/2^+ \rightarrow 1/2^-)$ 
will appear 
in $d{\rm B(E1)}/dE$ at $O(R_{\rm core}^2/R_{\rm halo}^2)$, and so the expansion 
of FSI and asymptotic-normalization effects in powers of the EFT small 
parameter  does not capture all of the physics of the B(E1) transition. 

We do, however, have the advantage that the {\it same} counterterm enters 
both the bound-to-bound and bound-to-continuum transition, so the number of 
free parameters needed to describe B(E1) strength is limited. Up to NLO 
accuracy for the bound-to-bound, and NNLO for the bound-to-continuum, there 
are five numbers: $\gamma_0$ and $\gamma_1$ (which are known from separation 
energies), and $r_1$, $A_0$, and the counterterm, which must be fitted to the 
available data on  B(E1) ($1/2^+ \rightarrow 1/2^-$) and $d{\rm B(E1)}/dE$. 

The quality of the Coulomb excitation data is such that realistically we 
can hope only to extract one parameter from it, and so an NNLO analysis is 
not feasible at this time. In the next section we present our numerical 
results from the NLO analysis of the Coulomb excitation data. At NLO we need 
only 
$\gamma_0$, $\gamma_1$, $r_1$, and $A_0$ (or, equivalently, $r_0$).

\section{Results for Coulomb excitation of the ${}^{11}$Be nucleus}

\label{sec-Couldisresults}

The data set we use is that of Ref.~\cite{Pa03}. The predictions obtained in 
the previous section must be folded with the neutron detector resolution as 
described in that paper. For the number of E1 photons as a function of  the photon energy $\omega$
we take~\cite{Be09}
\begin{equation}
N_{E1}(\omega)=\frac{2 Z^2 \alpha}{\pi v^2}\left[
\xi K_0\left(\xi \right) K_1\left(\xi \right) 
- \frac{v^2}{2} \xi^2
\left[K_1^2\left(\xi \right) - K_0^2\left(\xi \right)\right]\right]\,,
\end{equation}
where $\xi=\omega b/(\gamma v)$, with $b$ the impact parameter, and $\gamma=(1 - v^2/c^2)^{-1/2}$ the Lorentz factor of the beam.  
We adopt the value $b=10.38$ fm and obtain $\gamma$ from the kinetic energy 
of the ${}^{11}$Be beam of 520 MeV per nucleon, both based on values quoted 
in Ref.~\cite{Pa03}.

\begin{figure}[ht]
\centerline{\includegraphics[width=0.7\columnwidth]{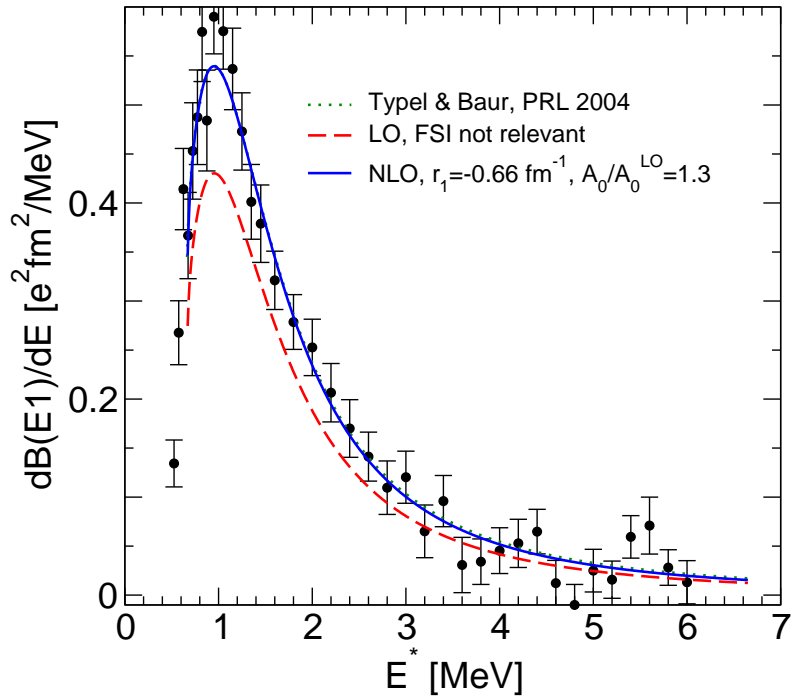}}
\caption{Differential B(E1) strength for Coulomb dissociation of Beryllium-11 
into a neutron and a ${}^{10}$Be nucleus, plotted versus the excess energy of 
the detected neutron $E^*$, in MeV. The data are from Ref.~\cite{Pa03}. The 
theory curves have been folded with detector resolution. The red dashed line 
is the leading-order Halo EFT prediction, which does not include any 
final-state interactions. Final-state interactions, with the effective range 
taking on a value fixed from the bound-to-bound E1 transition strength, are 
included in the NLO result, which is shown in blue. The result of 
Ref.~\cite{TB05} is the green dotted line, which essentially matches the 
solid blue line.}
\label{fig:results}       
\end{figure}
The leading-order result (\ref{eq:dBE1dELO}) can then be folded with detector 
response and the spectrum of E1 photons. This produces the red-dashed curve 
in Fig.~\ref{fig:results}. At next-to-leading order we take the value of 
$r_1$ fixed as per Eq.~(\ref{eq:r1LO}). We then have one free parameter, 
the value of $A_0$ at NLO (or, equivalently $r_0$)~\cite{Ph99}. 
After folding we find a reasonable fit for $A_0/A_0^{\rm LO}=1.3$, 
which corresponds to $r_0=2.7$ fm. The green 
dotted line shows the result of Typel and Baur of Ref.~\cite{TB05}. That 
result was obtained with all integrals regulated at a scale of $R=2.78$ fm. 
This corresponds to specific assumptions about all the counterterms that 
appear in the theory.

Note that $A_0/A_0^{\rm LO}=1.3$ increases the charge radius of the ${}^{11}$Be 
ground state to 
\begin{equation}
\langle r_E^2 \rangle^{(\sigma)}_{{}^{11}{\rm Be}}=2.42~{\rm fm},
\label{eq:NLOrcs}
\end{equation}
which is certainly in agreement with the atomic-physics number (\ref{eq:atphysnumber}) within the expected size of NNLO corrections. 
This must be taken with a grain of salt, though, since the change in the 
$A=10$ to $A=11$ radius difference from LO to NLO is $\sim 30$\%. But, in 
contrast to observables involving p-wave binding, the radius of the 
${}^{11}$Be ground state does not receive any corrections from 
short-distance physics until N$^3$LO. The remaining difference between 
our NLO number (\ref{eq:NLOrcs}) and the experimental number 
is certainly 
consistent with the presence of the short-distance operator 
$\sim L_{C0}^{(\sigma)}$ at N$^3$LO in the EFT expansion for the radius.

The neutron radius of the ground state of ${}^{11}$Be is similarly increased by the presence of an $r_0 > 0$. The value of $A_0$ extracted from the Coulomb dissociation data produces
$
\langle r_n^2 \rangle^{1/2}=5.6~{\rm fm}
$
within the context of our NLO calculation, a 30\% shift from the LO result. As with $\langle r_E^2 \rangle$, the next contribution is from purely short-distance physics, and assuming that this gives a contribution of order $R_{\rm core}$ to $\langle r_n^2 \rangle^{1/2}$, we have the Halo EFT determination:
\begin{equation}
\langle r_n^2 \rangle^{1/2}=(5.6 \pm 0.6)~{\rm fm},
\label{eq:rnNLO}
\end{equation}
where the error bar does not account for the statistical uncertainty in the extraction of $A_0$. Eq.~(\ref{eq:rnNLO}) 
is a model-independent result for the neutron radius of ${}^{11}$Be, obtained from the Coulomb dissociation data of Ref.~\cite{Pa03}. 

\section{Correlations between observables in the ${}^{11}$Be system}

Effective field theories in general, and Halo EFT in particular, provide 
model-independent correlations
between different observables. In the previous sections, we have 
expressed the electromagnetic properties of the ${}^{11}$Be system
through the effective range parameters for n-${}^{10}$Be scattering,
$\gamma_0$, $\gamma_1$, $r_0$, and $r_1$. Our expressions thus
can be interpreted as correlations between scattering observables and 
electromagnetic properties. However, depending on the
experimental information available, it could be useful to look at
correlations between electromagnetic observables.

As an example, we consider the correlation between the B(E1)
strength and the radius of the p-wave state in ${}^{11}$Be at LO.
Using Eqs.~(\ref{eq:LOpradresult}, \ref{eq:BE1}) we obtain
\begin{equation}
{\rm B(E1)}=\frac{2 e^2 Q_c^2}{15 \pi} \langle r_c^2 \rangle^{(\pi)}
x \left[
\frac{1+ 2 x}{(1+ x)^2}\right]^2\,,
\label{eq:BE1corr}
\end{equation}
where $x=\sqrt{B_1/B_0}$ is the ratio of the neutron separation energies
for the p-wave and s-wave states. The B(E1) strength is thus proportional 
to the mean-square radius of the p-wave state. 
In the limit of vanishing neutron
separation energy for the p-wave state, the B(E1) strength vanishes 
linearly with $x$.
Eq.~(\ref{eq:BE1corr}) can also be used to obtain the electric radius of the 
p-wave state $\langle r_E^2 \rangle^{(\pi)}$ directly from the measured value of
${\rm B(E1)}$ and the neutron separation energies $B_1$ and $B_0$.
This gives a radius of the $^{11}$Be p-wave state, relative to the $^{10}$Be
ground state:
\begin{equation}
\langle r_E^2 \rangle^{(\pi)} =0.35...0.39\mbox{ fm}^2\,,
\label{eq:Rpwave_corr}
\end{equation}
depending on whether the value for B(E1) from Eq.~(\ref{eq:CEXBE1})
or  Eq.~(\ref{eq:lifeBE1}) is used. 

Analogously, the strength for the E1 transition to 
the continuum can be related to the radius of the s-wave state. However, this observable is also affected by the 
magnitude of p-wave final-state interactions. At next-to-leading order we have:
\begin{equation}
\frac{d{\rm B(E1)}}{dE}^{\rm NLO}
= \frac{12}{\pi^2 B_0} \frac{y^3}{(1+y^2)^4}
\left[e^2 Q_c^2 \Delta\langle r^2_E\rangle^{(\sigma)}
     -\pi {\rm B(E1)} \frac{(1+x)^4 (1+3y^2)}{(y^2+x^2)(1+2x)^2}
\right]
\label{eq:dBE1dEcorr}
\end{equation}
where $y=p'/\gamma_0$. 
Similar correlations can be derived for other observables and/or
at higher orders. They have proven useful in the analysis of 
universal properties in ultracold atoms and also show promise 
for halo nuclei \cite{Braaten:2004rn}.

\section{Conclusion}

This discussion of electromagnetic observables in the Beryllium-11 system 
already displays the significant recent experimental activity that has been 
focused on this system. Coulomb excitation has been used at a variety of 
facilities to probe E1 transitions, and atomic-physics experiments have made 
great strides through advances in trapping technology. This means that the 
time is ripe for a detailed analysis of electromagnetic properties of halo 
systems. Here we have shown how EFT can provide such an analysis. 

The Halo EFT we employed is complementary to {\it ab initio} 
methods~\cite{Fo05,Fo09,QN08}, which can struggle to describe E1 transitions 
and radii in these extended nuclei because of the widely varying core and 
halo scales that are present in the problem. In Halo EFT this wide separation 
of scales is the basis for the calculation. Input that summarizes the physics 
at scale $R_{\rm core}$ can be taken from either simulation or experiment, and 
the EFT is then used to predict the outcome of experiments that probe dynamics 
at the halo scale. 

The results derived here in EFT mirror elegant analytic approaches to halo nuclei 
(see, e.g. Refs.~\cite{TB04,TB08,TB05}). But, in contrast to those works, 
there is no regulator dependence in our result for the E1 strength. This 
is a consequence of current conservation in our formalism. Moreover, the EFT
delineates the order at which any observable receives a contribution
from physics at scale $R_{\rm core}$---a contribution which cannot be
calculated using the asymptotic wave functions (\ref{eq:LOwfs}). 
The EFT also explicitly shows whether---and if so, where---these 
short-distance effects appear  in other observables. 
In this way we can systematically assess the impact of
physics at scale $R_{\rm core}$ on low-energy electromagnetic observables in 
halo nuclei, and so 
go beyond the calculations of Refs.~\cite{TB04,TB05,TB08}.

We determined the magnitude of the s-wave n-${}^{10}$Be effective range, 
$r_0$, by examining experimental 
results for the low-energy E1 strength function in breakup to the 
${}^{10}$Be-neutron channel: $d$B(E1)$/dE$. We find a reasonable fit for 
$r_0=2.7$ fm. We were also able 
to extract a value for the p-wave effective range $r_1$ from the 
bound-to-bound B(E1) strength, obtaining 
$r_1=-0.66$ fm$^{-1}$ up to the 40\% corrections which seem typical in this 
EFT. This gives a p-wave scattering volume $a_1=(374 \pm 150$) fm$^3$. This 
overlaps the range quoted in Ref.~\cite{TB05}, but has a lower central value. 
We would caution against any strong conclusion regarding values of $a_1$ and
$r_1$  until the counterterm that parameterizes the short-distance 
piece of the E1 strength in this system has been determined by further 
measurements. 

There are no spectroscopic factors in our approach. 
Nevertheless, a comparison of the values we have obtained for the physical 
observables $r_0$, $r_1$ and $a_1$ (or, equivalently, $A_0$ and $A_1$) with 
those found within many-body models and {\it ab initio} calculations would 
be interesting. While the interpretation of these quantities in terms of the 
many-body dynamics of the $A=11$ system is {\it not} one of the goals of 
this work, such an interpretation could inform which observables are 
interesting for future investigation within the EFT, and, in particular, 
which ones will most reward study at higher EFT orders.

In this vein, we note that a higher-order calculation of the E1 observables 
considered here would necessitate determination of the E1 counterterm that 
occurs at NLO in the bound-to-bound B(E1) strength. Better data on Coulomb 
dissociation of ${}^{11}$Be at low energies would be very helpful in this 
regard, since the same counterterm occurs in the continuum E1 strength at 
next-to-next-to-leading order. With this one additional parameter we should 
be able to perform calculations to 5\% or better for $d{\rm B(E1)}/dE$. It is, 
though, important to note that the EFT used here breaks down at energies of 
order a few MeV, and so the window in which the theory will give such an 
accurate description of data is quite limited. 

But, already at the order to which we have worked here,
the extraction of parameters for n-${}^{10}$Be scattering facilitates 
predictions for the electric and neutron radii of ${}^{11}$Be. We find 
agreement at the level expected of our NLO calculation between our result 
for $\langle r_E^2 \rangle^{1/2}_{{}^{11}{\rm Be}}$ and the measurement of 
Ref.~\cite{No09}. We also predict that the excited state of ${}^{11}$Be
has an electric radius of $2.43 \pm 0.1$ fm. Meanwhile, the neutron radius 
of the ${}^{11}$Be ground state is found to be 
$\langle r_n^2 \rangle^{1/2}=5.6 \pm 0.6$ fm. We note that EFT provides a 
way to extract this quantity from the Coulomb dissociation data which is 
independent of any assumptions about the details of the physics that occurs 
at distances $\sim R_{\rm halo}$. 

Lastly, we emphasize that the Halo EFT can be used to derive correlations
driven by the separation of scales in the system. These correlations may not be
obvious in {\it ab initio} approaches where this separation is not 
explicit in the parameters of the theory. EFT has been successful at revealing these
patterns in few-nucleon systems, e.g. the correlation between the triton binding energy
and the spin-doublet neutron-deuteron scattering length (the \lq\lq Phillips
line''~\cite{Phillips68}), and the correlation between the triton and alpha
particle binding energies (the \lq\lq 
Tjon line''~\cite{Tjo75}). We now understand that those correlations are 
driven by the hierarchy between NN scattering lengths and the range of the NN force: $1/a_s, 1/a_t \ll m_\pi$, and
so they can both be derived 
within an appropriate EFT~\cite{EfT85,Pl_alpha}.
Here we have presented analogous correlations for electromagnetic observables
in the $^{11}$Be system. Similar relations have proven of significant interest 
in the field of ultracold atoms
 \cite{Braaten:2004rn}. Future tests of correlations such as 
Eqs.~(\ref{eq:Rpwave_corr}) and (\ref{eq:dBE1dEcorr}) can reveal the extent 
to which the physics of ${}^{11}$Be, and that of other halo nuclei, is 
driven by ``universality". 
 
The application of the Halo EFT to other one-neutron 
halos, and to two-neutron halos such as ${}^{11}$Li, are obvious next steps.
With the inclusion of Coulomb interactions, proton halos become 
accessible as well.

\section*{Acknowledgments}
We are grateful to S.~Typel for supplying us with the data of Ref.~\cite{Pa03} 
and for assistance with the details of the convolution of theoretical results 
with detector resolution needed to obtain Fig.~\ref{fig:results}.
We thank L.~Platter for discussions regarding the Clebsch-Gordon formalism used here for the spin-1/2 and spin-3/2 fields, 
and C.~Ji for pointing out Ref.~\cite{Sak} to us. 
This research was supported by the US Department of Energy under grant 
DE-FG02-93ER40756, by the BMBF under contract
06BN9006, and by the Mercator programme of the Deutsche 
Forschungsgemeinschaft. We thank the US Institute of Nuclear Theory for 
its support during the program ``QCD, Cold Atoms, and Few-hadron Systems", 
which fostered this collaboration. 
DRP thanks the HISKP for its hospitality, and its members for a very enjoyable 
time in Bonn. 


\end{document}